\documentclass[draft]{agujournal2019}
\usepackage{url} 
\usepackage{lineno}
\usepackage[inline]{trackchanges} 
\usepackage{soul}
\usepackage[utf8]{inputenc}
\usepackage{amssymb}
\usepackage{amsmath}
\usepackage{mathrsfs}

\draftfalse

\journalname{JGR: Planets}

\begin{document}

\title{Investigating Thermal Contrasts Between Jupiter's Belts, Zones, and Polar Vortices with VLT/VISIR}

\authors{
Deborah Bardet\affil{1}, 
Padraig T. Donnelly\affil{2}, 
Leigh N. Fletcher\affil{1}, 
Arrate Antuñano\affil{3}, 
Michael T. Roman\affil{1}, 
James A. Sinclair\affil{4}, 
Glenn S. Orton\affil{4}, 
Chihiro Tao\affil{5},
John H. Rogers\affil{6}, 
Henrik Melin\affil{1}, 
Jake Harkett\affil{1}}

\affiliation{1}{\emph{School of Physics and Astronomy, University of Leicester}, address: University Road, Leicester LE1 7RH, United Kingdom}
\affiliation{2}{\emph{Laboratoire de M\'{e}t\'{e}orologie Dynamique / Institut Pierre-Simon Laplace (LMD/IPSL), Sorbonne Universit\'{e}, Centre National de la Recherche Scientifique (CNRS), \'{E}cole Polytechnique, Institut Polytechnique de Paris, \'{E}cole Normale Sup\'{e}rieure (ENS), PSL Research University}, address: Campus Pierre et Marie Curie BC99, 4 place Jussieu, 75005 Paris, France}
\affiliation{3}{\emph{UPV/EHU, Escuela Ingernieria de Bilbao, Fisica Aplicada}, address: Spain}
\affiliation{4}{\emph{Jet Propulsion Laboratory/ California Institute of Technology}, address: 4800 Oak Grove Drive, Pasadena, CA 91109, USA}
\affiliation{5}{\emph{National Institute of Information and Communications Technology (NICT)}, address:
4-2-1, Nukui-Kitamachi, Koganei, Tokyo 184-8795, Japan}
\affiliation{6}{\emph{British Astronomical Association}, address: London, United Kingdom}

\correspondingauthor{Deborah Bardet}{db528@leicester.ac.uk}

\begin{keypoints}
\item Jupiter’s belt/zone thermal structure extends from the equator to the edge of the polar vortices and correlates with the zonal jets.
\item The cold polar vortices coincides with reflective aerosols, suggesting dynamical entrainment by the northernmost and southernmost jets.
\item Temporal evolution of the southern auroral hotspot shows a stratospheric cooling after a solar wind compression event.
\end{keypoints}

\begin{abstract}
Using images at multiple mid-infrared wavelengths, acquired in May 2018 using the VISIR instrument on ESO's Very Large Telescope (VLT), we study Jupiter's pole-to-pole thermal, chemical and aerosol structure in the troposphere and stratosphere. 
We confirm that the pattern of cool and cloudy anticyclonic zones and warm cloud-free cyclonic belts persists throughout the mid-latitudes, up to the polar boundaries, and evidence a strong correlation with the vertical maximum windshear and the locations of Jupiter's zonal jets.
At high latitudes, VISIR images reveal a large region of mid-infrared cooling poleward $\sim$64$^{\circ}$N and$\sim$67$^{\circ}$S extending from the upper troposphere to the stratosphere, co-located with the reflective aerosols observed by JunoCam, and suggesting that aerosols play a key role in the radiative cooling at the poles.
Comparison of zonal-mean thermal properties and high-resolution visible imaging from Juno allows us to study the variability of atmospheric properties as a function of altitude and jet boundaries, particularly in the cold southern polar vortex. 
However, the southern stratospheric polar vortex is partly masked by a warm mid-infrared signature of the aurora. 
Co-located with the southern main auroral oval, this warming results from the auroral precipitation and/or joule heating which heat the atmosphere and thus cause a significant stratospheric emission.
This high emission results from a large enhancement of both ethane and acetylene in the polar region, reinforcing the evidence of enhanced ion-related chemistry in Jupiter’s auroral regions.

\end{abstract}
\section*{Plain Language Summary}
As NASA's Juno spacecraft lacks instrumentation sensitive to the mid-infrared, we have used the ground-based Very Large Telescope Imager and Spectrometer (VISIR) on ESO's Very Large Telescope (VLT) to support the mission since 2016. 
The May 2018 dataset provides a comprehensive view of Jupiter's pole-to-pole thermal, chemical, and aerosol structure; including the Great Red Spot; and the auroral-related heating in the southern polar stratosphere.
We show that the alternating pattern of cool cloudy zones and warm cloud-free belts correlate to the zonal winds locations, extends from the equator to the boundary of the polar vortices ($\sim$64$^{\circ}$N and$\sim$67$^{\circ}$S).
Comparing this ground-based dataset with Juno observations, we show that the polar regions experience radiative cooling from the upper troposphere to the stratosphere, probably induced by the reflective polar aerosols, and confined at the poles by the polar vortices.
However, the southern stratospheric polar vortex is partly masked by a large auroral emission, a signature of an interaction between the magnetosphere and stratosphere of Jupiter, which causes a $\sim20$ K warming as deep as the middle stratosphere. 
Excess hydrocarbons within the polar domain reinforces the evidence of enhanced ion-related chemistry in Jupiter’s auroral regions, resulting from charged particles precipitating from the top of the atmosphere to lower altitudes.

\section{Introduction}
\label{sec:Introduction}

The meridional distribution of chemical trace species in the middle atmosphere of Jupiter provide crucial insights into the thermal and dynamical processes occurring in this planetary atmosphere. 
Despite being observed for decades, Jupiter remains enigmatic in terms of its global circulation, eddy-to-mean interaction from the troposphere to the upper atmosphere, chemical processes and the meridional and vertical transport of chemicals. 

For instance, the latitudinal distribution of chemical species such as ammonia \cite{Gier:86,Acht:06,dePa:16Jupiter_clouds,Li:17Jupiter_ammonia}, phosphine \cite{Flet:09,Gile:17Jupiter_phosphine,Gras:20}
and para-hydrogen \cite{Conr:98,Flet:17Jup_paraH2}, combined with the observed temperature and aerosol distributions, may suggest that Jupiter's banded appearance of belts and zones is related to meridional circulation patterns associated with each zonal jet, and that the sense of the circulation changes dramatically as a function of height in Jupiter's troposphere \cite{Flet:21}.
This proposed ``stacked-cell'' circulation may be comprised of deep Ferrel-like cells below the clouds, dominated by eddy forcing of zonal winds \cite{Duer:21}, and upper cells above the clouds of eddy dissipation and zonal wind decay with altitude \cite{Inge:20,Show:05,Flet:20}.
The vertical shear on the zonal winds may be inferred from the meridional temperature gradients via the thermal wind equation \cite{Holt:04}, and previous studies have shown that the zonal winds appear to decay with altitude from the cloud top to the tropopause \cite{Pirr:81,Conr:90}. 
The dissipation mechanism, and its source, remains unclear and has never been directly observed, but could result from wave or vortex stresses opposing the winds \cite{Pirr:89,Orso:93Jupiterlinearmodel}.

These meridional circulation cells could explain the differences in temperatures, aerosols, and gaseous composition between the cold, cloudy anticyclonic zones and the warm, comparatively cloud-free cyclonic belts. 
In the upper troposphere, zones experience adiabatic expansion above the clouds and below the stably stratified tropopause, and are favourable to condensation of ices that enhance aerosol reflectivity; whereas belts experience adiabatic compression, explaining their comparatively cloud-free structure.
Moreover, zones are surrounded by eastward-propagating jets at their polar edges, and westward-propagating jets on their equatorward edges. 
Conversely, the belts exhibit westward jets along their poleward boundaries and eastward jets on their equatorward boundaries.
Strong alternating eastward- and westward-propagating zonal jets generate potential vorticity gradients that act as barriers to meridional mixing \cite{Read:06}. 
Observations from Voyager/IRIS and Cassini/CIRS suggest that this banded pattern of temperature, aerosol and zonal winds persists up to high latitudes near 60$^{\circ}$ in the upper troposphere \cite{Conr:83,Flas:86,Simo:06,Flet:16jupiter}, but neither spacecraft provided good coverage of the mid-infrared brightness gradients at Jupiter's high latitudes in the polar domains. 
To explore Jupiter's polar regions in the mid-infrared, we require the high spatial resolution offered by world-class ground-based observatories with primary mirror diameters of 8~m or larger to mitigate the blurring effects of diffraction.  

Jupiter’s stratosphere shows signs of equally rich dynamics as its troposphere, with numerous wave signatures in the temperature fields and isolated thermal disturbances in tropical and auroral regions \cite{Sinc:17aurora_hydrocarbons_PartI,Flet:17Jup_NEB,Gile:23}.
As with the circulations in the troposphere, stratospheric circulations are mostly deduced from the observation of anomalies in the distribution of trace species, such as hydrocarbons.
The main by-products of methane photochemistry, ethane (C$_2$H$_6$) and acetylene (C$_2$H$_2$), display surprising differences in their meridional distribution in the middle stratosphere (from 1 to 10~mbar).
While acetylene displays a maximum at low latitudes, following the yearly-averaged insolation as is expected from photochemistry, long-lived ethane increases towards the poles up to the $\sim$70$^{\circ}$--boundary \cite{Nixo:07,Nixo:10,Zhan:13,Flet:16jupiter,Meli:18,Giles:21Jupiter_meridC2H2}.
In the same vein, HCN and CO$_2$, introduced into Jupiter's stratosphere following comet Shoemaker–Levy 9 (SL 9) impact in 1994, follow different migration schemes: HCN has been efficiently mixed from the impact site (44$^{\circ}$S) to northern mid-latitudes, while CO$_2$ has been greatly enhanced near the south pole \cite{More:03,Grif:04,Lell:06,Cava:17,Sinc:23}.
Photochemical models, including parameterizations of meridional, vertical diffusion and advection, attempted to address the observed opposite distributions of C$_2$H$_6$ and C$_2$H$_2$, or HCN and CO$_2$, but none could satisfactorily reproduce the observations \cite{Hue:18,Lell:06,Zhan:13}.
To date, there is no clear picture of the global dynamical transport and the chemical processes on those chemical species during their journey from the equator toward the polar latitudes. 

However, poleward of the $\sim$70$^{\circ}$ boundary, observations of Jupiter's atmosphere reveal interactions between its magnetosphere and stratosphere, in particular through the polar aurora interaction with the middle atmosphere. 
Auroral regions at Jupiter's northern and southern poles are prone to injections of high-energy ions and electrons coming from the jovian magnetosphere and solar wind compression events.
As they propagate downwards into the neutral atmosphere, those high-energy charged particles deposit their energy down to approximately 1 mbar \cite{Flas:04}, resulting in a magnetosphere-to-stratosphere warming (caused by the charged particle precipitation, ion drag and Joule heating).  
Moreover, the increased electron flux in auroral regions also leads to increased rates of ion-neutral reactions and electron recombination, thus altering the chemistry in auroral regions compared to non-auroral regions and lower latitudes \cite{Sinc:17aurora_hydrocarbons_PartI, Sinc:18aurora_hydrocarbons_PartII, Sinc:19aurora_hydrocarbons_PartIII}.
The increased flux of electrons in the auroral regions is sensed in both the mid-infrared emission and ultra-violet absorption: (i) by modifying the thermal and chemical structure of the atmosphere, there is an enhancement of the mid-infrared emission chemical species in the stratosphere, such as methane (CH$_4$), acetylene (C$_2$H$_2$), ethylene (C$_2$H$_4$) and ethane (C$_2$H$_6$); (ii) the modification of the chemical structure can result in an increase in abundance, that acts to enhance the absorption in the ultra-violet \cite{Cald:80,Cald:83,Dros:93,Kim:85,Kost:93,Live:93, Sinc:17aurora_hydrocarbons_PartI, Sinc:18aurora_hydrocarbons_PartII, Sinc:19aurora_hydrocarbons_PartIII}.

Despite mid-infrared observations over several decades, the interaction between auroral particle penetration and the vertical and horizontal chemistry in the deeper layers of the neutral atmosphere remain poorly understood.
Indeed, as auroral processes are the result of charged particles precipitating from the top of the atmosphere to higher pressures, any effect of the aurora in the lower layers is expected to be a continuous top-down effect.
\citeA{Sinc:17aurora_hydrocarbons_PartI} detected two discrete vertical levels of heating. 
On one hand, they found heating at 10~$\mu$bar, corresponding to the lower boundary of Jupiter's thermosphere during an auroral event, plausibly due by chemical heating, H$_2$ dissociation from excited states and Joule heating from Pedersen currents \cite{Grod:01,Yate:14,Badm:15}.  
On the other hand, they also found a deeper layer of heating at 1~mbar, which could be due to adiabatic heating by subsidence in the auroral hotspot regions \cite{Sinc:17aurora_hydrocarbons_PartI}, whose horizontal diffusion would be confined to the pole by an asymmetric auroral jet surrounding the main auroral oval.  
In fact, ground-based observations of stratospheric polar HCN and CO highlighted strong westward and eastward jets at 0.1~mbar, respectively located at 70$^{\circ}$S and 55$^{\circ}$N, delimiting the main auroral emission boundaries \cite{Rego:99}, enclosing within the main auroral oval the subsidence heating from the sub-$\mu$bar levels \cite{Chau:11} toward the middle stratosphere \cite{Cava:21} through adiabatic compression. 
Such auroral jets could act as a guide to confine aurora-heated, hydrocarbon-enriched gas to the pole during the entire advection of the gas to deeper pressures, before being diffused or transported toward other latitudinal and/or longitudinal regions outside of the main auroral oval.
Moreover, the polar C$_2$H$_2$ (from the auroral ion-neutral chemistry) is suspected to be transported outside the auroral oval, toward quiescent polar regions and toward equatorward regions, where the neutral chemistry dominates and where acetylene may be quickly converted into ethane \cite{Sinc:17aurora_hydrocarbons_PartI,Gile:23}. 
This hypothetical transport and transformation of acetylene into ethane out of the auroral main oval might explain the homogenisation of C$_2$H$_6$ across the entire region \cite{Sinc:17aurora_hydrocarbons_PartI}, as well as the poleward enhancement of ethane observed from space and ground-based facilities \cite{Nixo:07,Flet:16jupiter,Meli:18} that remains unexplained by neutral photochemical models \cite{Hue:18}. 

This paper presents new measurements of the spatial variability of temperature, aerosols, and gaseous species associated with Jupiter's banded structure, polar vortex boundary, auroral heating, and large-scale features like the Great Red Spot.  
We use a global mid-infrared map of Jupiter acquired by the VISIR instrument (Very Large Telescope Imager and Spectrometer for InfraRed) on ESO's Very Large Telescope (VLT) in May 2018.  
The high spatial resolution afforded by VLT allows us to measure infrared contrasts at Jupiter's mid-latitudes and polar vortices that were inaccessible to flyby spacecraft (Voyager and Cassini), and beyond the reach of observations from 3-m diameter facilities like the IRTF \cite{Flet:16jupiter} due to more severe diffraction of the beam.  
The May-2018 global map was the most complete (in terms of spatial and spectral coverage) dataset acquired as part of a 2016-2022 programme of support for NASA's Juno spacecraft. 
This dataset was obtained simultaneously with Juno's 13$^{th}$ perijove (closest approach), and provides an unprecedented view of Jupiter's poles in the mid-infrared, allowing us to compare simultaneous Juno observations to the mid-infrared maps.

The dataset and the reduction and calibration methods are presented in section \ref{sec:obs_processing_method}. 
Different methods of spectral retrieval are explored in section \ref{sec:retrieval}. 
We describe the retrieval results, with a focus on the belt/zone thermal structure and chemical species distribution, and their implication on the jovian atmospheric dynamics in section \ref{sec:belts_zones}. 
Section \ref{sec:GRS} provides insights about the thermal and cloud structures of the Great Red Spot.
The polar cloud distribution, with comparison to Juno observations on Perijove 13, will be provided in section \ref{sec:polar_regions}, and the time variation of the auroral-related thermal signature in the stratosphere will be inferred in section \ref{sec:polar_aurora_retrieval}.
Finally, we will draw conclusions and perspectives in section \ref{sec:conclusion}.

\section{Observations and processing method}
\label{sec:obs_processing_method}

\subsection{VLT/VISIR 2018 May 24$^{th}$-27$^{th}$ dataset} 
As NASA's Juno spacecraft lacks instrumentation sensitive to the mid-infrared (7-25 $\mu$m), a programme of ground-based infrared support has been underway since 2016 \cite <i.e.>[]{Flet:17Jup_NEB,Flet:18Jup_mesowaves,Antu:20,Antu:21,Flet:20Jupiter_equatorialplume}. 
To this purpose, we use the VLT Imager and Spectrometer for the mid-infrared (VISIR, \citeA{Laga:04}) on the ``unit telescope'' UT3 Melipal.
The array of VISIR is 1024$\times$1024 pixels, with a pixel scale of 0.045'' providing a field of view of 38$\times$38''.
The maximum chop throw of 25'' combined with the maximum angular size of Jupiter of $\sim$44'' requires independent observations to acquire the northern and southern hemisphere, sometimes spread over several nights.
To observe an entire hemisphere in all wavelengths, it is necessary to observe multiple images in a sequential way, in a block of approximately 50 minutes that is then repeated as Jupiter rotates.
The instrument provides diffraction-limited filtered imaging at high sensitivity in the M
band (at 4.82 $\mu$m), N band (8-13 $\mu$m) and Q band (16-25 $\mu$m) of the mid-IR, chopping and nodding to remove the sky background.
It also features a long-slit spectrometer that performs N-band spectroscopy with spectral resolutions varying between 150-30,000 (not used in this study).

\begin{figure}
    \centering
    \includegraphics[width=0.9\textwidth]{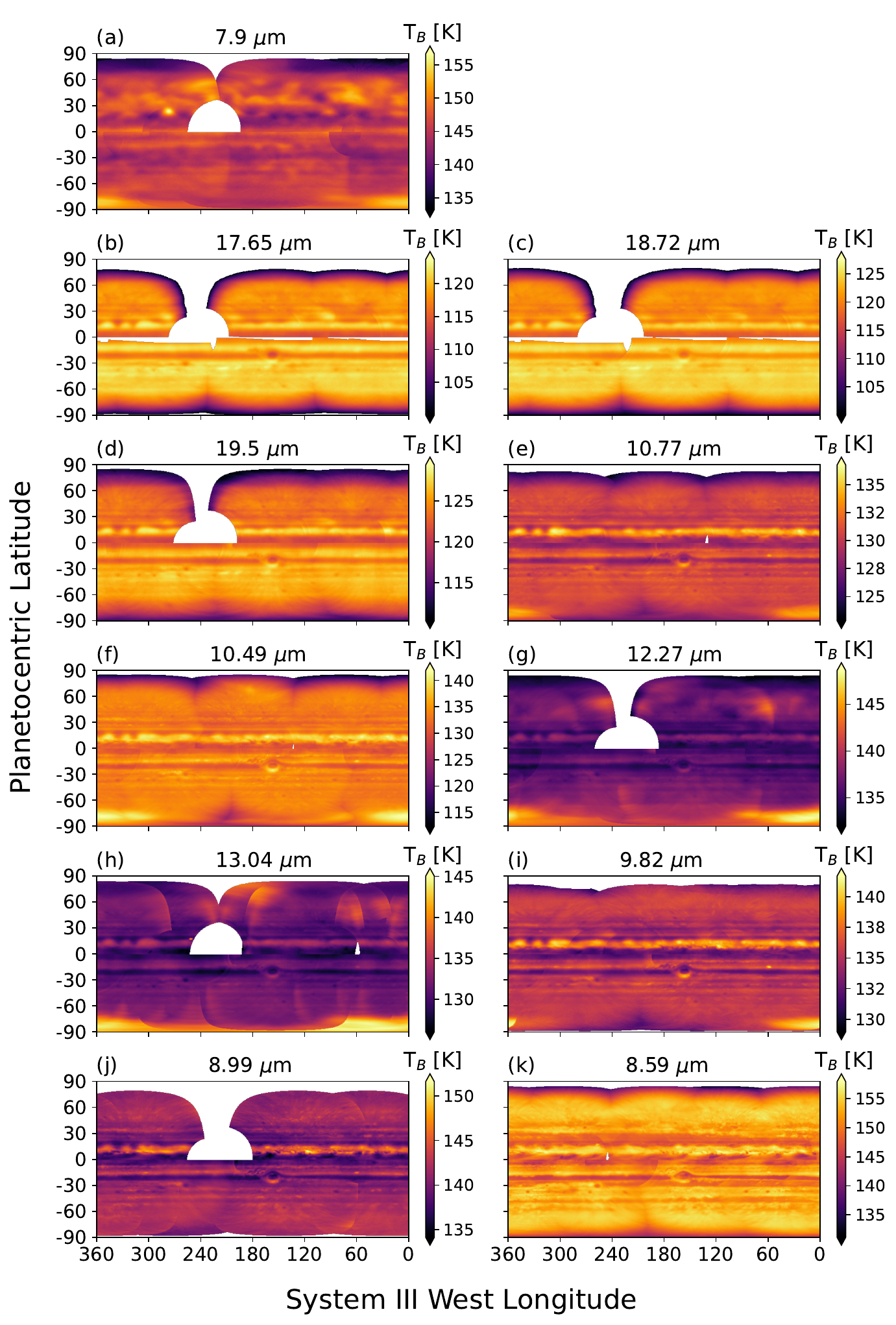}
    \caption{Global maps for the VLT/VISIR N-Band (7.9 to 13.04$\mu$m) and Q-Band (19.50, 18.72 and 17.65 $\mu$m) filters between 2018 May 24$^{th}$ and 27$^{th}$, obtained by combining the individual calibrated cylindrical maps. Maps shown here have been also corrected to remove the limb-darkening or limb-brightening effect (see section \ref{subsec:limb_correction}). Gaps near 240$^\circ$W are due to missing data}
    \label{fig:global_maps}
\end{figure}

Images at multiple wavelengths between 4.9 and 19.5~$\mu$m have been analysed to study the thermal, chemical and aerosol structure of Jupiter's belts, zones, and polar domains. 
In particular, a ``visitor mode'' observing run in May 2018 (coinciding with Juno's perijove 13) provided global coverage of Jupiter in eleven narrow-band filters (Figure \ref{fig:global_maps}).
This dataset is composed of 281 observations, taken on the 24$^{th}$ May 2018 (02:29–07:45 UT), 24$^{th}$-25$^{th}$ May 2018 (23:02–04:43 UT), 25$^{th}$-26$^{th}$ May 2018 (23:06–08:01 UT) and 26$^{th}$-27$^{th}$ May 2018 (22:46–04:41 UT).
This 11-mid-infrared-filter set mainly senses the stratospheric temperature through the methane (CH$_4$) band emission (7.9~$\mu$m), 
tropospheric temperature via the collision-induced hydrogen-helium continuum (13.04, 17.65, 18.72, 19.50~$\mu$m),
temperature and aerosol opacity deeper in the troposphere (8.59 and 8.99~$\mu$m), tropospheric distribution of ammonia gas and temperature via NH$_3$ absorption bands (8.59, 8.99, 9.82, 10.49, 10.77 and 12.27~$\mu$m), 
acetylene (C$_2$H$_2$, 13.04~$\mu$m) and ethane (C$_2$H$_6$, 12.27~$\mu$m). 
The latter two wavelengths are particularly interesting in the polar regions, as the contribution of the hydrocarbon emissions become as strong or stronger than the contribution from the continuum emissions. 
Thus, those two filters are more sensitive to the hydrocarbon distribution at high emission angles than at low emission angle, implying a higher sensitivity to the hydrocarbons in the polar regions than at the equator. 
In addition, there are two other filters that sense the NH$_3$ and temperatures at 11.25 and 11.88~$\mu$m; but these wavelengths were only used once to image the southern hemisphere, so are not included in our global analysis.
For this dataset, Jupiter's atmosphere is covered from 89$^{\circ}$S to 87$^{\circ}$N (in planetocentric latitude), and describes a sub-observer of -3.76$^{\circ}$ (offering a better view of the southern pole). 
To determine the maximum latitude up to which we trust the observations, we use the diffraction limit, define as 1.22$\lambda$/D, where $\lambda$ is the considered wavelength and D the mirror diameter of the telescope. 
For the longest wavelength of this dataset (19.5 $\mu$m), the diffraction limit is $\sim$0.6 arcsec. 
For simplicity, we consider Jupiter as a sphere with an angular radius at the observational time of 22.2 arcsec. 
In the case of optimal condition of observation, the farthest north we can point the telescope and keep the beam fully on Jupiter is half the diffraction limit at 19.5 $\mu$m (i.e. 0.3 arcsec). 
Hence, we estimate that within a disk of 0.3 arcsec for the longest wavelength of the dataset, the maximum latitude is $\sim$80$^{\circ}$ (i.e., $lat_{max}= arcsin(\frac{22.2 - 0.3}{22.2})$), for a perfect sphere and a sub-observer latitude of 0$^{\circ}$.
Hence, for any measurements within 10$^{\circ}$ plus the sub-observer latitude from either pole at 19.5$\mu$m, the planetary flux is be convolved with the much colder deep sky flux.  
Therefore, poleward of 83$^{\circ}$S and 77$^{\circ}$N, Jupiter's thermal flux can be considered as contaminated for the purposes of the thermal retrieval.
In the following sections, we are not taking into account the data poleward of those two latitude thresholds. 
We also note that hereafter all latitudes will be expressed in planetocentric form.

The eleven global maps obtained after image processing and correction (explained in subsection \ref{subsec:data_processing}) permit an overview of some of Jupiter's famous features, which will be detailed in the present study, such as the banded structure of cool zones and warm belts, and large vortices such as the Great Red Spot and Oval BA.  
Jupiter's equatorial stratospheric oscillation \cite<sometimes called the quasi-quadrennial oscillation ``QQO'',>[]{Leov:91} was in a phase with a warm equatorial band at 7.9 $\mu$m in 2018.

\begin{figure}
    \centering
    \includegraphics[width=\textwidth]{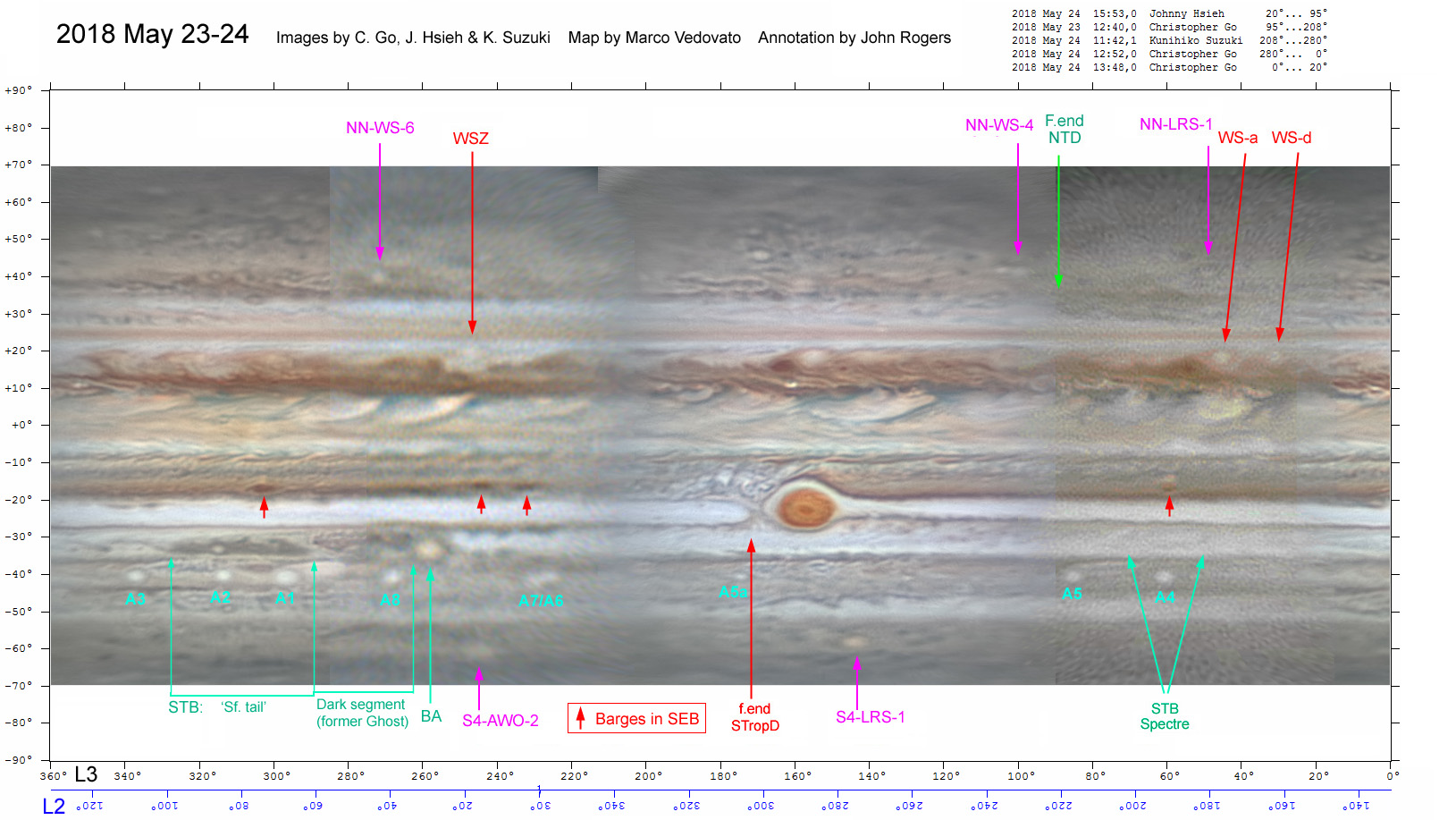}
    \caption{Ground-based amateur observations during the 2018 May 23$^{rd}$ to 24$^{th}$ night taken by Christopher Go, Johnny Hsieh and Kuknhiko Suzuki, processed and mapped by Marco Vedovato and annotated by John Rogers. Published on the British Astronomical Association, into the ``Jupiter in 2018, Report no.5'' \url{https://britastro.org/section_information_/jupiter-section-overview/jupiter-in-2017-18/jupiter-in-2018-report-no-5}}
    \label{fig:global_map_amateur}
\end{figure}

Several bright spots are depicted in these global maps.
For instance, the bright spot seen in the 60$^{\circ}$W region of the Southern Equatorial Belt (SEB) in the 10.77-, 10.49-, 9.82-, 8.99-, and slightly in the 8.59-$\mu$m filters, which seems correlated with one of the SEB Barges labelled in the ground-based amateur observations map (Figure \ref{fig:global_map_amateur}).
These cyclonic barges are bright in the mid-infrared and dark in the visible due to some combination of warmer temperature, aerosol clearing, and depleted NH$_3$.
In the stratosphere and at the tropopause level (i.e., at 7.9-, 17.65-, 18.72-$\mu$m on Figures \ref{fig:global_maps}a-\ref{fig:global_maps}c, respectively), there is an intense bright spot located at 280$^{\circ}$W/25$^{\circ}$N, reminiscent of the 14-K-warm anomaly observed from May 28–29 to June 5 in 2017 at 178$^{\circ}$W/28$^{\circ}$N at a pressure of 1.2 mbar, but disappeared from July 2017 to February 2018 \cite{Gile:20}.
\citeA {Gile:20} showed that this stratospheric anomaly occurred at the same time as a disruption of the QQO, similar to Saturn's equatorial oscillation disturbance by the stratospheric ``beacon'', resulting from the Great White Spot in 2010-2011 \cite{Flet:17}. 
Additional study is needed to confirm whether this 280$^{\circ}$W/25$^{\circ}$N bright spot observed with VISIR is linked to a new disruption event of the downward propagation of the QQO.

\begin{figure}
    \centering
    \includegraphics[width=0.5\textwidth]{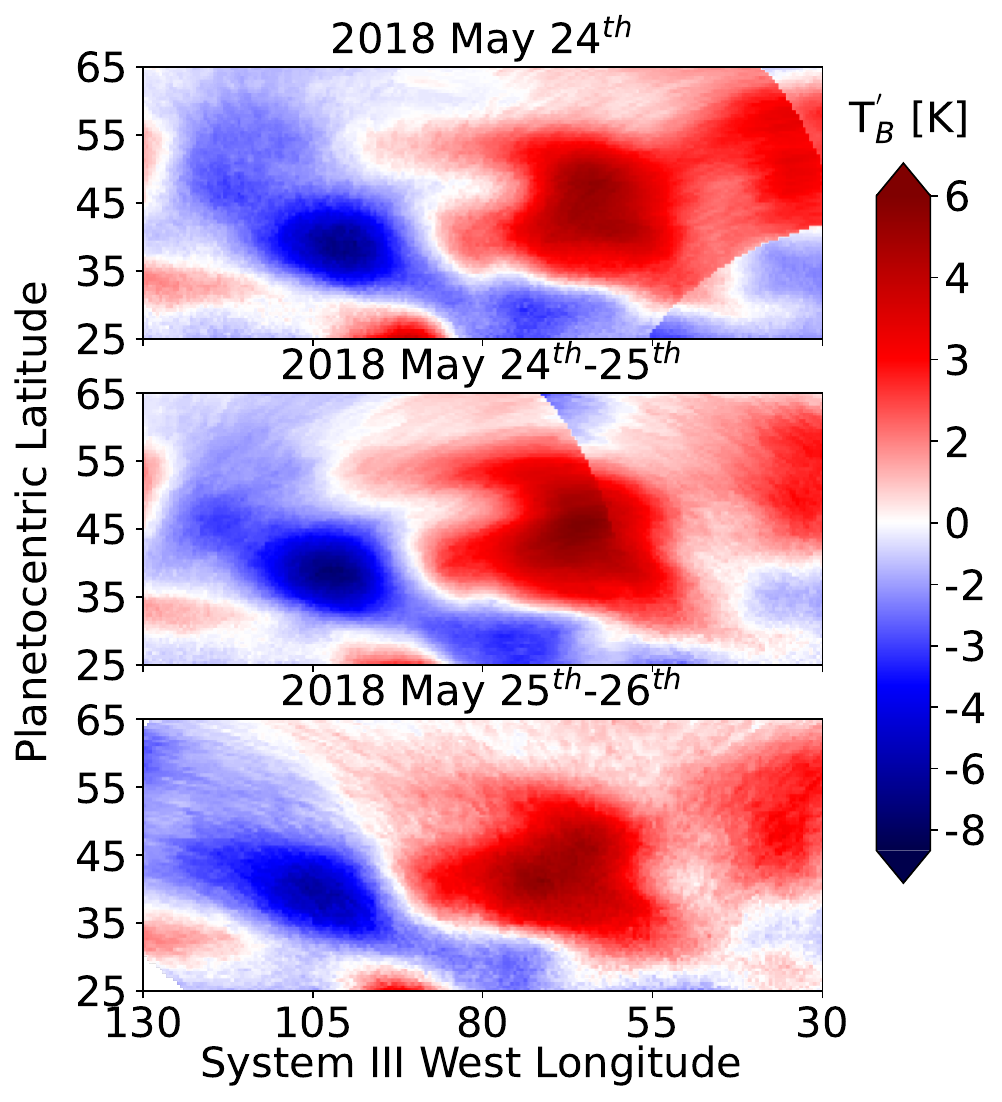}
    \caption{Details of zonal brightness temperature perturbation maps, calculated from the zonally averaged profile of each global map per night and for each filter to highlight the thermal perturbation at high northern latitudes in the stratosphere. Only the first three nights of observations are shown here, because this part of the atmosphere was not observed during the last night.}
    \label{fig:Rossby_waves_patterns_TB_perturbation}
\end{figure}

The 7.9-$\mu$m filter reveals significant brightness temperature variability in the northern stratosphere, suggesting large wave activity during Jupiter's northern winter (the season on Jupiter in 2018 May), in contrast with the southern hemisphere showing a more quiescent brightness temperature behavior.
In particular, in the 50-60$^{\circ}$N latitude band, there are two warm anomalies (in terms of T$_B$) respectively located at 70 and 230$^{\circ}$W System III (Figure \ref{fig:global_maps}a), also seen in the 12.27-$\mu$m filter sensing C$_2$H$_6$. 
By subtracting the zonal-mean brightness temperature, we can see the anomaly clearly in Figure \ref{fig:Rossby_waves_patterns_TB_perturbation} over multiple nights (only the 70$^{\circ}$W anomaly is showed on Figure \ref{fig:Rossby_waves_patterns_TB_perturbation}). 
Both of these warm anomalies appear to be located on the eastward edge of a cold anomaly.  
The persistence of these two repetitive features near $50^\circ$N, equidistant from each other, is reminiscent of a planetary-scale Rossby wave, comparable to those circumpolar waves observed around the southern pole \cite{BarrIzag:08}. 
Due to the short timescale of the present dataset, it is not possible to highlight any preferential direction of propagation and, thus determine if the cold anomaly propagates to equatorward latitudes through time, characteristic of a baroclinic system.

\begin{figure}
    \centering
    \includegraphics[width=\textwidth]{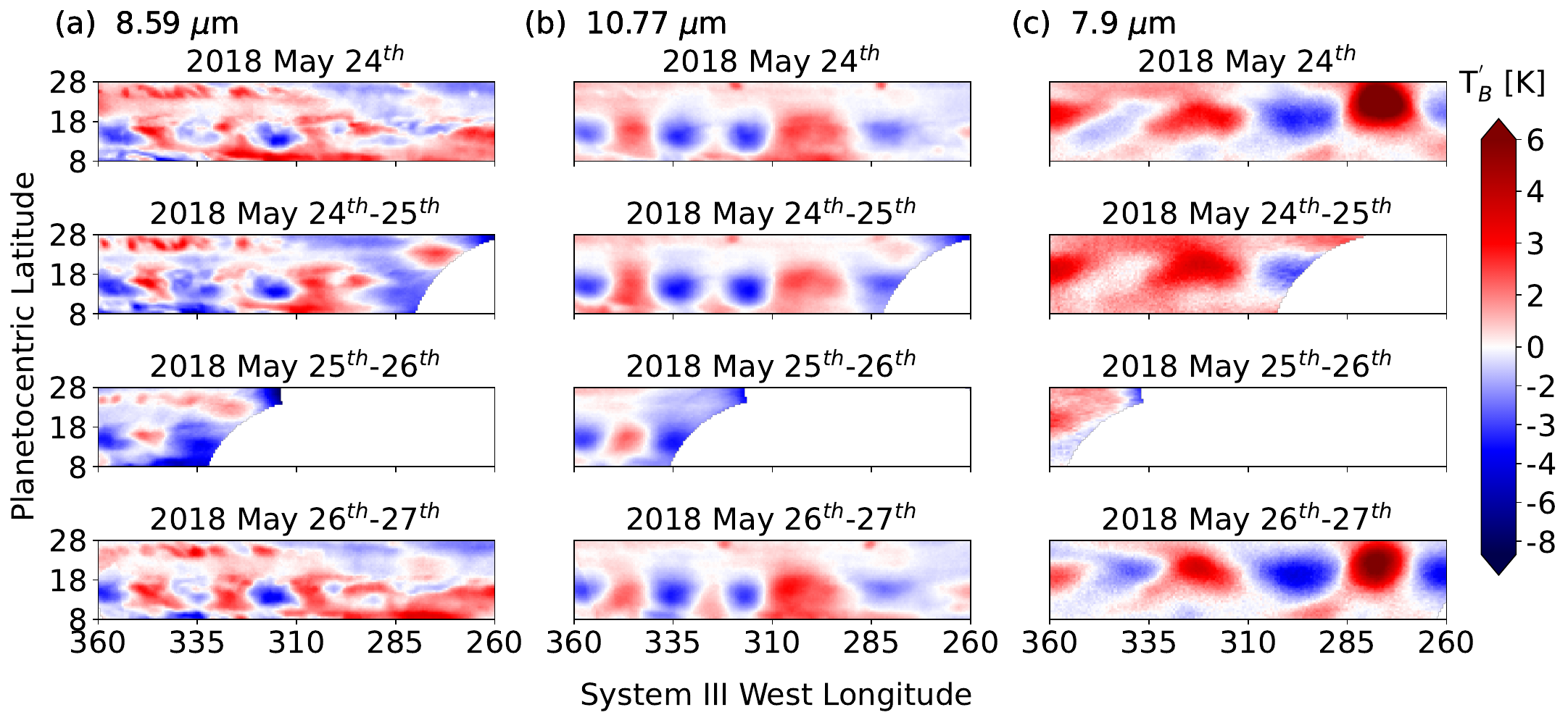}
    \caption{Details of zonal brightness temperature perturbation maps, calculated from the zonally averaged profile of each global maps per night and for each filter to highlight the thermal perturbation induce by any planetary-scale waves in the Northern Equatorial Belt (NEB) and the Northern Tropical Belt (NTB). Here, only the maps for the (a) 8.59, (b) 10.77 and (c) 7.90~$\mu$m filters are shown as they depict the larger signal of the wave-induced thermal perturbation for all nights.}
    \label{fig:waves_patterns_TB_perturbation}
\end{figure}

Two other wave patterns are visible around the northern tropical regions of Jupiter, the first one in the tropospheric Northern Equatorial Belt (NEB) centered near the 15$^{\circ}$N latitude, and the second one in the stratospheric Northern Tropical Belt (NTB) centered at 20$^{\circ}$N, both reminiscent of the two slow-propagating thermal waves observed in 2015/2016 during the NEB expansion with VLT/VISIR \cite{Flet:17Jup_NEB}.
Between 10 and 25$^{\circ}$N, a pattern of consecutive warm and cold anomalies from 360 to 300$^{\circ}$W is observed in all tropospheric filters, extended up to the tropopause, and observed for the four nights of observation (Figures \ref{fig:waves_patterns_TB_perturbation}a and \ref{fig:waves_patterns_TB_perturbation}b).
In the same vein, temperature perturbation maps of the stratospheric 7.9$\mu$m-filter depict a feature of consecutive cold and warm anomalies from 360 to 260$^{\circ}$W.
In both cases, the zonal-mean brightness temperature perturbations maps, focusing on the areas where the waves are most visible, show some time variations in the shape and extension for numerous thermal anomalies, however there is no significant displacement of the patterns toward the east or the west during this 4-day window.  
These waves appear stationary within the four days of the observations, consistent with the slow propagation of NEB and NTB waves observed by \cite{Flet:17Jup_NEB}.
Hence, the determination of the physical characteristics of these waves, such as the wave speed, direction of propagation, etc., remain unfeasible for the short time series of these data.  

Most of the upper-tropospheric filters (i.e., wavelengths larger than 10~$\mu$m), as well as the stratospheric 7.9-$\mu$m filter, display a well-defined cold polar vortex at each pole (Figure \ref{fig:global_maps}), in particular outside of the main auroral oval and/or for filters where the auroral-related emission is not too bright to mask the cold vortex boundaries. 
For instance, the three Q-band filters (19.50, 18.72 and 17.65 $\mu$m) show a significant cooling from $\sim$65$^{\circ}$ to the pole in both hemispheres. 
Because of the large mid-infrared emission in the southern auroral oval (360-300$^{\circ}$W and 90-0$^{\circ}$W System III), the southern polar vortex is not readily visible in the N-band maps. 
However, the north polar vortex boundary is easily seen in the stratosphere (7.9-$\mu$m) and in the acetylene and methane emission band (12.27 and 13.04-$\mu$m) as a significant cold area all around the pole, only disturbed by the plausible (but weaker than the southern one) northern auroral-related brighter emission (in the northern aurora oval, from 260$^{\circ}$ to 150$^{\circ}$W System III).   
A detailed characterisation of the polar dynamics, the size of the polar vortices, and the auroral-related heating will be provided in sections \ref{sec:polar_regions} and \ref{sec:polar_aurora_retrieval}.

\subsection{VLT/VISIR data processing}
\label{subsec:data_processing}

Reduction of VLT/VISIR observations followed standard techniques for mid-infrared datasets \cite{Flet:09} and are briefly summarised here.  

\subsubsection{Detector readout noise}
When illuminated by a bright source like Jupiter, VISIR displays vertical and horizontal fixed-pattern noise that must be removed before data analysis.  
Vertical striping noise is the result of differences in detector readout noise and inhomogeneous illumination by the extended source of Jupiter. 
The multiplicative nature of detector noise in the thermal infrared leads to a flux-dependent striping pattern which cannot be easily removed by flat-fielding or dark-frame subtraction. 
Planetary and deep-space pixels are separated by a flux threshold to approximate the location of the limb, and band-pass filtering is performed on each detector half independently to avoid mixing readout noise. 
A combination of low-pass and high-pass filtering (vertical column-averaging then horizontal Gaussian smoothing) is used to isolate the striping pattern for subtraction from planetary pixels.
The central burst of noise is the dominant horizontal component of the fixed pattern noise and is treated analogously to the vertical component. 
\citeA{21donnelly} elaborates on this filtering technique and uses quantitative validation metrics to show that, although noise is never perfectly removed, jovian information is optimally preserved.

\subsubsection{Geometric registration} 
The chop-nod subtracted and destriped images have been processed, assigning latitude, longitude and emission angle values to each location on the disc. 
To do this, orbital ephemerides of the planet at the time of each observation were used to generate a virtual image of the planetary limb that is manually fitted to the observed limb by eye. 
The observed radiance was then projected into a stereographic latitude-longitude map (0.5$^{\circ}$ × 0.5$^{\circ}$ resolution), along with the emission angles, the radial Doppler velocity, and the solar incidence angle for each latitude and longitude. 
This limb-fitting method has been used for other mid-IR imaging datasets and was estimated to produce errors no larger than $\pm$0.2~K in brightness temperature and 1.5$^{\circ}$ in latitude \cite{Flet:09}.

\subsubsection{Radiometric calibration \& uncertainties}
\label{subsubsec:calib_profiles}
\begin{figure}
    \centering
    \includegraphics[width=\textwidth]{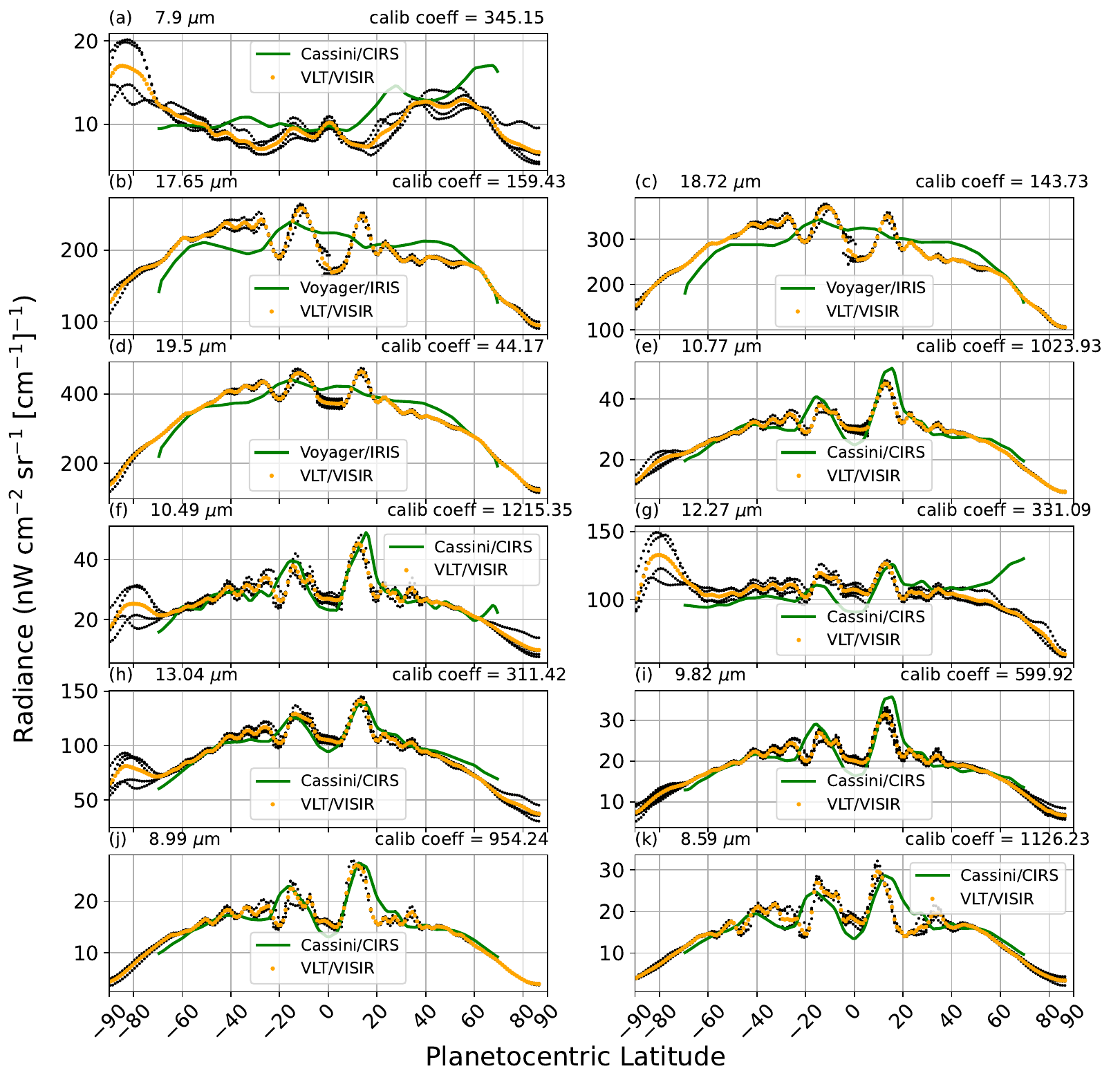}
    \caption{Calibrated central-meridian radiances (orange) for the VLT/VISIR N-Band (7.9 to 13.04 $\mu$m) and Q-Band (19.50, 18.72 and 17.65 $\mu$m) filters using Cassini/CIRS and Voyager/IRIS meridional profiles (green), respectively.  Individual central meridian measurements are shown as black dots. The calibration coefficient of the pole-to-pole radiance profiles (defined as the faction between the mean value of the radiance profile observed by VLT/VISIR and the mean value of the radiance profile observed by Cassini/CIRS or Voyager/IRIS) is indicated on the top right of each panel. }
    \label{fig:calib_profile}
\end{figure}

Each cylindrical map was initially radiometrically calibrated using the technique outlined in \citeA{Flet:09}. 
For each VLT/VISIR filter, a central-meridian zonal-mean profile of radiance was extracted for each observation, using a box of 60$^{\circ}$-longitude width, centered at the central meridian longitude (in System III, LCMIII) for each map. 
Hence, we have selected from 6 to 8 individual central meridian profiles per filter, which were averaged together to obtain a pole-to-pole meridian profile of radiance per VLT/VISIR filter.  
Finally, the averaged pole-to-pole meridian profiles were calibrated to a low-latitude (from 5$^{\circ}$S to 5$^{\circ}$N for the 7.9$\mu$m filter) or a full averaged meridional profile of radiance from spacecraft data (Figure \ref{fig:calib_profile}).  
N-band measurements (7.90 to 13.04~$\mu$m) were scaled to the zonal-mean radiance profile from the Cassini Composite Infrared Spectrometer (CIRS), captured in December 2000 -- January 2001 from a near-equatorial flyby, minimising the emission angle effects for the low latitudes (Figures \ref{fig:calib_profile}a and \ref{fig:calib_profile}e to \ref{fig:calib_profile}k). 
However, because the Cassini/CIRS Q-band measurements were not resolved over latitude, the VLT/VISIR Q-band measurements (17.65, 18.72 and 19.5~$\mu$m,  respectively in Figures \ref{fig:calib_profile}b, \ref{fig:calib_profile}c and \ref{fig:calib_profile}d) were scaled to the zonal mean radiance from the Voyager InfraRed Interferometer \& Spectrometer (IRIS) \cite{Demi:89, Orto:94, Demi:97, Orto:98, Simo:06}. 
The zonally-averaged Cassini/CIRS and Voyager/IRIS meridional radiance profiles were convolved with the VLT/VISIR filter functions and the telluric transmission spectrum (calculated with ATRAN, \citeA{Lord:92}).
\newline

This method of radiometric scaling assumes that the average global temperature of Jupiter has not changed systematically over time, leaving the dataset insensitive to changes between the Voyager and Cassini observations, as noted by \citeA{Orto:91, Simo:07}. 
A global systematic change in mean temperature could cause a systematic difference in the retrieved temperatures, although \citeA{Simo:06} show the zonal mean structure of Jupiter measured by Voyager/IRIS and Cassini/CIRS and the global mean temperature did not appear to have changed between the datasets (1979-2000).  
Additionally, a comparison of Voyager data and data from the Stratospheric Observatory for Infrared Astronomy (SOFIA) by \citeA{Flet:17Jup_paraH2} suggests a thermal change of no larger than 5~K in the Q band over the 35-year period. 
Nevertheless, the infrared brightness of localised regions of Jupiter is known to vary dramatically - the shifting pattern of the stratospheric oscillation; the fades, revivals, and upheavals of the equatorial belts and zones; and interhemispheric changes that may be related to seasonal variability \cite<e.g.,>[]{Antu:20,Orto:22}.  
By scaling the VISIR data to spacecraft observations across all latitudes within $\pm60^\circ$ of the equator, our radiometric calibration is insensitive to these localised brightness changes. 
\citeA{Antu:20} estimated the error introduced by this scaling to be 0.95 K in the worst case, less than 0.05 K in the best case, and 0.3 K on average.  
In summary, assuming negligible global temperature changes provides a consistent calibration across this VLT/VISIR dataset used hereinafter.

In addition to the systematic uncertainty described above, the measured VLT/VISIR radiances are subject to random uncertainties associated with the electrical and thermal state of the electrons in the VISIR CCDs (dark current), and the inherent variability of the telluric atmosphere. 
The high quality of the VISIR AQUARIUS sensors means that the majority of the variability in pixel brightness is real, so we can estimate the random noise by the standard deviation of the background sky pixels in each image. 
This is typically less than 5\% of the mean radiance in each image across all filters, so as a conservative estimate we apply a consistent 5\% uncertainty to all radiance points in each spectrum. 
The resulting uncertainty from random noise throughout the N and Q bands is 0.4-0.9 K in brightness temperature \cite{Antu:20, Antu:21}.

\subsection{Centre-to-Limb Corrections}
\label{subsec:limb_correction}
\begin{figure}
    \centering
    \includegraphics[width=\textwidth]{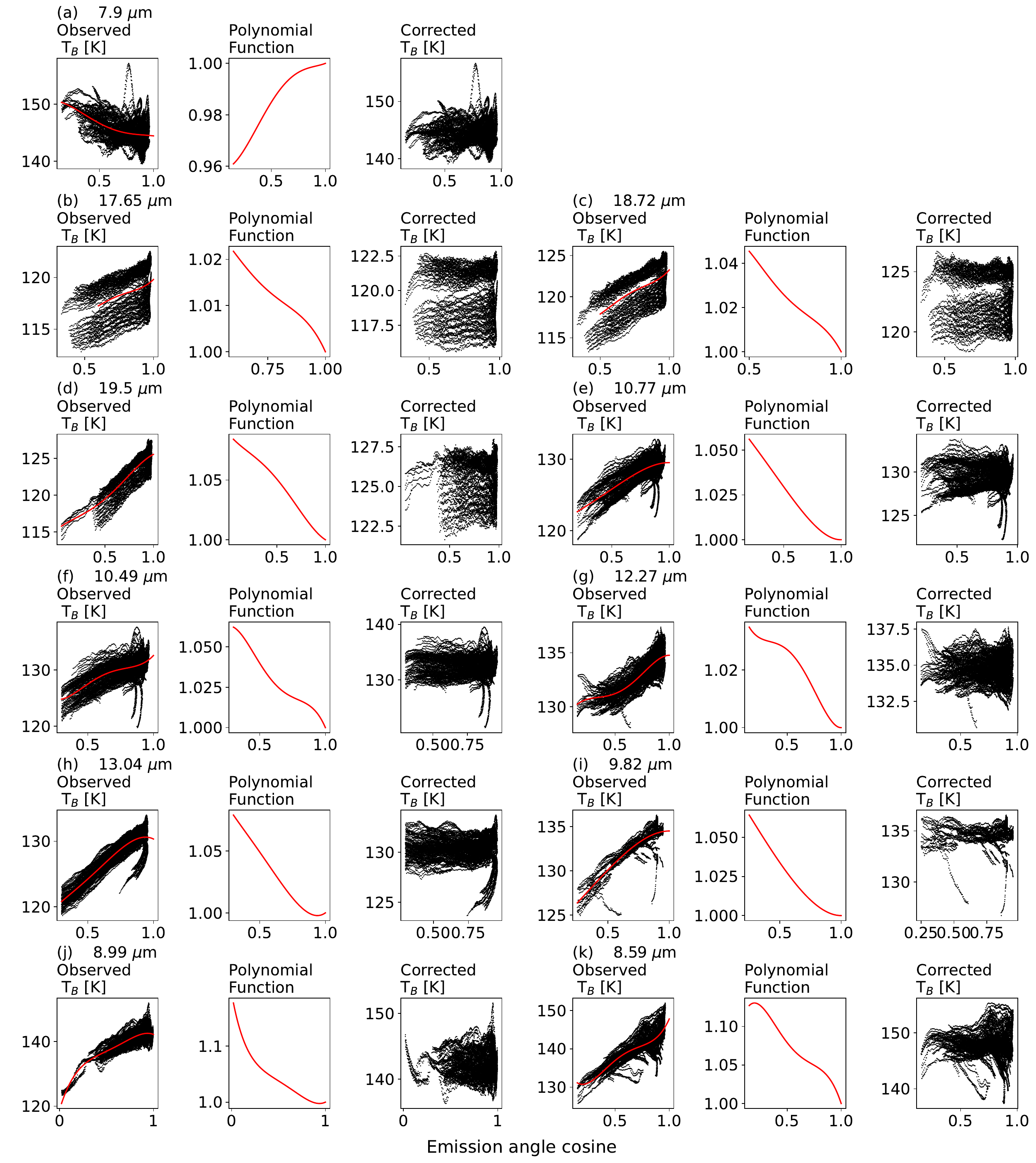}
    \caption{Polynomial adjustment of VLT/VISIR N-Band (a and e-k) and Q-Band (b, c and d) to correct the centre-to-limb darkening/brightening effects to enable merging of individual observations into a cylindrical map. For each filter, the left panel shows the brightness temperature variations over the selected latitude band (black dots) superimposed with the corresponding fourth order polynomial function (red line); the middle panel shows the variations of the centre-to-limb factor correction ($p(1)/p(\mu)$ where $\mu$ is the cosine of the emission angle); and the right panel show the corrected brightness temperature variations over the selected latitude band, i.e., $T_B \times p(1)/p(\mu)$. These three fields are shown as a function of the cosine of the emission angle.}
    \label{fig:polynome_adj}
\end{figure}

For the purposes of displaying the multiple VISIR observations as a global map, we need to remove the limb darkening (troposphere-sensing filters) or limb brightening (stratosphere-sensing filters) from each observation.
For each filter, we calculated an average variation of brightness temperature as a function of the cosine of the emission angle, and corrected each cylindrical map with this adjustment.
For the 10 N-Band filters (the stratospheric 7.9~$\mu$m and the tropospheric 8.59 to 13.04~$\mu$m filters), we averaged the two tropical latitude bands from the brightness temperature maps, from 5 to 25$^{\circ}$ of latitude in each hemisphere, and calculated a 4th-order polynomial function of the brightness temperature variation in terms of emission angle (Figures \ref{fig:polynome_adj}a and \ref{fig:polynome_adj}e to \ref{fig:polynome_adj}k).
For the three Q-Band filters (17.65, 18.72, 19.50~$\mu$m), we only used the southern tropical band of brightness. 
Effectively, several tests using both the northern and the southern tropical band for the Q-Band filters failed in the correction of the centre-to-limb darkening effect. 
The only way to correct these three filters was to remove the Northern Equatorial Belt (NEB) in the sample of data used to calculate the polynomial adjustment (Figures \ref{fig:polynome_adj}b to \ref{fig:polynome_adj}d). 
Hence, we obtained an unique polynomial adjustment function for each VLT/VISIR filter, that was applied to each cylindrical map to correct the limb effects, and afterward those corrected cylindrical maps were combined to construct the global maps shown in Figure \ref{fig:global_maps}.  
Of course, as we use an average centre-to-limb polynomial adjustment, some maps still remain a reduced limb-brightening or limb-darkening effect (for instance in the northern polar regions of the 10.77, 10.49, 9.82 or even 8.99-$\mu$m filters) that does not prevent us from identifying polar phenomena in the following sections.

\section{Retrieval}
\label{sec:retrieval}

\subsection{Retrieval calculation pipeline}

To derive temperatures, aerosols, and gaseous abundances from the VLT/VISIR 2018 May 24$^{th}$-27$^{th}$ dataset, we used the Non-linear optimal Estimator for MultivariatE spectral analySIS (NEMESIS, \citeA{Irwi:08}) retrieval model, comprised of a radiative transfer code and an inversion algorithm.

NEMESIS uses the radiative transfer equations to forward-model mid-infrared spectra from a reference atmosphere. 
This reference atmosphere is described by several prior profiles of temperature, aerosols and molecular species from previous observations and modelling studies \cite{Roma:93,Seif:98,Niem:98,Mose:05jupiter,Nixo:07,Flet:09}.
The reference atmosphere contains vertical distributions of temperature, ammonia (NH$_3$), phosphine (PH$_3$), acetylene (C$_2$H$_2$) and ethane (C$_2$H$_2$) from a 30$^{\circ}$N-to-30$^{\circ}$S meridian average of Cassini/CIRS results, gridded from 10 bar to 1 $\mu$bar on 120 vertical levels \cite{Nixo:07,Flet:09}.
The temperature profile derived from Cassini/CIRS originally used the $T(p)$ from the Galileo probe Atmospheric Structure Instrument (ASI, \citeA{Seif:98}) as prior.
Diacetylene (C$_4$H$_2$) and ethylene (C$_2$H$_4$) prior profiles are included based on the results from the photochemical model of \citeA{Roma:96}. 
The methane (CH$_4$) prior profile results from the photochemical model calculations of \citeA{Roma:93}, reflecting the atmosphere based on the descent of the Galileo Probe \cite{Seif:98}, with some adjustments: the deep CH$_4$ abundance has been enhanced by 5\%, to match the results of the Galileo probe Neutral Mass Spectrometer experiment \cite{Niem:98}, and is decreasing in abundance at altitudes above the 10-$\mu$bar level, following the prediction from the photochemical model of \citeA{Mose:05jupiter}. 
The sources of spectral line data and collision-induced absorptions are identical to those described in \citeA{Flet:18hexagon}.  
The \emph{a priori} profile of aerosol is a single layer of ammonia ice, with particle radii of 10$\pm$5$\mu$m, between 800 and 100-mbar pressure levels with a scale height 0.5 times the gas scale height.

Observational input files for NEMESIS are the pole-to-pole meridian profiles of radiance (Figure \ref{fig:calib_profile}), which were used to calibrate the individual cylindrical maps as described in section \ref{sec:obs_processing_method}. 
We remind the reader that for this stage of the observations analysis, the cylindrical maps retain the limb-darkening (or limb-brightening for the 7.9-$\mu$m filter), in order to ensure the consistency with the radiance and the emission angle values of each bin for the retrieval calculations.
For this purpose, at each latitude point covered by VLT/VISIR, the average value of radiance (calculated to create the pole-to-pole meridian profiles of radiance displayed in orange on Figure \ref{fig:calib_profile}) is associated to the minimum value of the emission angle across the individual profiles (black dots on Figure \ref{fig:calib_profile}), in order to assign the pole-to-pole meridional profile to the most central view of the overall dataset at that particular latitude point, and to effectively remove longitudinal limb effects.
However, a global limb-darkening (or limb-brightening) effect remains when we move toward the pole, corresponding of the highest emission angle view. 
This poleward high emission angle is taken into account in the retrieval calculation.

NEMESIS uses a non-linear optimal estimation model (and a Levenburg-Marquardt iterative scheme) to adjust the reference atmosphere in order to reduce the differences between the forward-modelled and observed spectra, the process of minimising the cost function \cite{Rodg:00, Irwi:08}. 
The cost function represents the relative weight of the assumed atmospheric priors and measurements and is contained in the calculation of the chi-squared distribution (described in subsection \ref{subsec:retrieval_sensitivity}).

\subsection{Vertical Sensitivity}

\begin{figure}
    \centering
    \includegraphics[width=\textwidth]{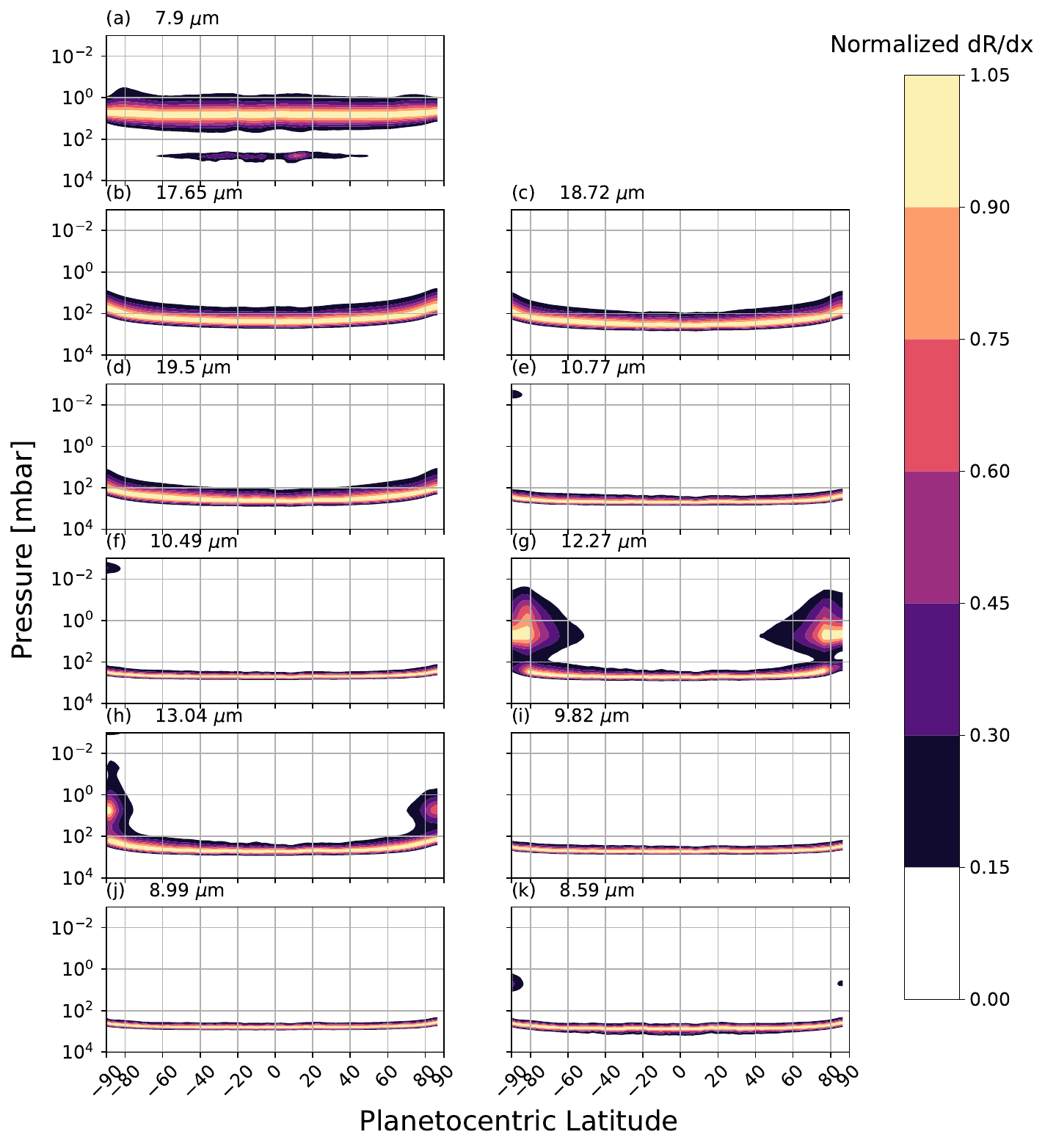}
    \caption{Meridional variations of the normalized contribution functions of Jupiter calculated from the VLT/VISIR N-Band ((a): 7.9 and (e-k): 10.77 to 8.59 $\mu$m) and Q-Band ((b-d): 19.50, 18.72 and 17.65 $\mu$m) filters set based on the varying emission angle.}
    \label{fig:weighting_functions}
\end{figure}

As expected, meridional variations of the contribution functions of the VISIR filters (Figure \ref{fig:weighting_functions}) reveal a displacement of the peak sensitivity towards lower pressure with increasing latitude (and therefore emission angle) for each filter. 
This is particularly evident in the Q-Band filters (17.65, 18.72 and 19.50~$\mu$m filters) sensing closest to the tropopause. 
However, the 7.90-, 12.27- and 13.04~$\mu$m filter contribution functions show several significant regions peaking at two distinct pressure levels, due to the increasing influence of stratospheric emission at higher emission angles. 
Hydrocarbons ethane and acetylene contribute to the 12.27 and 13.04 $\mu$m contribution functions at high emission angles, particularly poleward of 70$^{\circ}$, with contributions peaking at 10~mbar.
At lower latitudes, these hydrocarbons provide only a secondary contribution compared to the tropospheric continuum, which is why the maps in Figure \ref{fig:global_maps} show tropospheric structure at low latitudes, but stratospheric contributions at high latitudes. 
In addition, both 10.49 and 10.77-$\mu$m filters sense stratospheric ethylene poleward of 80$^{\circ}$S, for pressures lowers than 0.01~mbar.  
Preliminary retrieval tests, using the scaling approach (see following subsection) for ethylene, produced a flat retrieved meridional profile for this species up to $\pm$60$^{\circ}$ and a large increase peaking at $\pm$80$^{\circ}$. 
Those preliminary tests showed a lack of sensitivity of ethylene up to the polar boundaries for this dataset, as suggested by the flat 10.49 and 10.77-$\mu$m filter contributions (Figures \ref{fig:weighting_functions}f and \ref{fig:weighting_functions}e, respectively) for the whole latitude range. 
Hence, to take into account the polar enhancement, we have applied a scaling factor of 18 to the ethylene prior profile of \citeA{Roma:96}, creating a new ethylene prior profile that fits the VISIR data at all latitudes.
In the following retrieval runs, this new ethylene profile is held fixed. 
By fitting all of the VISIR filters simultaneously, NEMESIS is able to find the best-fitting model to these tropospheric and stratospheric contributions.

\subsection{Retrieval sensitivity to hydrocarbons and ammonia}
\label{subsec:retrieval_sensitivity}

Given that temperatures, gases, and aerosols are defined on 120 levels from 10 bars to 1 $\mu$bar, this implies a large number of degrees of freedom, and so a large number of variable parameters. 
Temperature inversions allow the temperatures to vary at every level, albeit with a smoothing assumption imposed by the \textit{a priori} uncertainty to remove non-physical oscillations in retrieved temperatures. 
Retrieved gases or aerosols either use a scaling of the model profile, or parameterise the vertical structure in some way. 
Several sensitivity tests (summarized in table \ref{tab:meridian_retrieval_tests}) have been performed to assess the best set of retrieval parameters to reproduce the VISIR observations. 
\begin{table}
    \caption{Summary of the conditions of each of the retrieval tests carried for this study.}
    \begin{center}
        \resizebox{\textwidth}{!}{
        \begin{tabular}{l c c c c c }
            \hline
            Retrieval name             & Temperature   & Aerosol profile & \multicolumn{2}{c}{Hydrocarbons profiles} & Ammonia profile\\
                                       &               &                 & C$_2$H$_2$ &  C$_2$H$_6$                  & NH$_3$\\
            \hline
            Temp                                     & retrieved     & -          & -          & -            & -   \\
            Temp Aer                                 & retrieved     & scaling    & -          & -            & -   \\
            Temp Aer C$_2$H$_2$ C$_2$H$_6$           & retrieved     & scaling    & scaling    & scaling      & -   \\
            Temp Aer NH$_{3}$                        & retrieved     & scaling    & -          & -            & scaling \\
            Temp Aer C$_2$H$_2$ C$_2$H$_6$ NH$_{3}$  & retrieved     & scaling    & scaling    & scaling      & scaling \\
            Temp Aer C$_2$H$_2$ C$_2$H$_{6, P}$      & retrieved     & scaling    & scaling    & parametric   & -   \\
            Temp Aer C$_2$H$_{2, P}$ C$_2$H$_{6}$    & retrieved     & scaling    & parametric & scaling      & -   \\
            Temp Aer C$_2$H$_{2, P}$ C$_2$H$_{6, P}$ & retrieved     & scaling    & parametric & parametric   & -   \\
            Temp Aer C$_2$H$_{2, P}$ C$_2$H$_{6, P}$ NH$_{3}$  & retrieved     & scaling   & parametric & parametric  & scaling   \\
            Temp Aer C$_2$H$_2$ C$_2$H$_6$ NH$_{3, P}$     & retrieved     & scaling   & scaling & scaling     & parametric \\
            Temp Aer NH$_{3, P}$                      & retrieved     & scaling   & -          & -            & parametric \\
            Temp Aer C$_2$H$_{2, P}$ C$_2$H$_{6, P}$ NH$_{3, P}$ & retrieved     & scaling   & parametric  & parametric & parametric \\
            \hline
        \end{tabular}
        }
    \end{center}
    \label{tab:meridian_retrieval_tests}
\end{table}

The simplest test, fixing the profiles of all gases and aerosols to be uniform with latitude and allowing only temperature to vary, proved incapable of reproducing the VISIR data within the uncertainties. 
Therefore, we have included aerosols and chemical composition as needed, and we have tested the sensitivity of the inversions to the choice of hydrocarbon and ammonia profiles.
In Table \ref{tab:meridian_retrieval_tests}, two different approaches to retrievals of chemical species have been used, for both, the fitting calculation is made at each latitudinal point assuring a latitudinal variation of chemical species distribution.
The ``scaling'' method allows the retrieval model to increase or decrease the entire profile, while retrieving a continuous temperature profile, in order to better fit the observations. 
This gives NEMESIS an extra degree of freedom per scaled gas, because both the temperature and the chosen parameter are varying. 
However, this ``scaling'' method is assuming a perfect knowledge of the vertical profile, i.e., a perfect knowledge about the vertical gradient of this profile, the position of any layer(s) of maximum abundance, and the maximum volume mixing ratio of this chemical parameter, etc.

To increase the degrees of freedom of the NEMESIS calculations, allowing more flexibility in chemical species profiles, we have explored the effect of using ``parametric'' retrievals of C$_2$H$_2$, C$_2$H$_6$ and NH$_3$ vertical profiles (see Table \ref{tab:meridian_retrieval_tests} for a summary of retrieval tests). 
This is especially needed at high latitudes, where prior measurements of the vertical distributions are more sparse \cite{Zhan:13,Giles:21Jupiter_meridC2H2}, and the low-latitude priors are less representative. 
In this configuration, the retrieval model is able to modify the maximum volume mixing ratio (i.e., mole fraction) of the well-mixed layer of a certain chemical specie, the transition pressure level (referred to as the knee of the profile), and the fractional scale height of the profile. 
Compared to the ``scaling'' method, the ``parametric'' method represents more freedom because chemical profiles change at every altitude, instead of all altitudes by the same amount.
For the VLT/VISIR 2018 May dataset, we have chosen to parameterize 
(a) an acetylene distribution with a constant volume mixing ratio of 1.02$\pm$0.8$\times$10$^{-5}$ for pressures lower than 0.05~mbar (according \citeA{Nixo:07} observations), and a decline with increasing pressure with a fractional scale height of 100$\pm$10; 
(b) an ethane distribution with a constant volume mixing ratio of 1.62$\pm$0.8$\times$10$^{-5}$ (according \citeA{Nixo:07} observations) for pressure p $<$ 0.2 mbar, and a decline with pressure with a fractional scale height of 2.0$\pm$1.0; and finally 
(c) an ammonia distribution with a constant volume mixing ratio of 3.62$\pm$0.4$\times$10$^{-4}$ (consistent with Juno observations of \citeA{Li:17Jupiter_ammonia}), for pressures higher than 800$\pm$200~mbar, and a decline with altitude with a fractional scale height of 0.15$\pm$1.0.
Because of the simple shape of the vertical profiles of C$_2$H$_2$, C$_2$H$_6$ and NH$_3$ when the ``parametric'' retrieval is used, the true vertical distribution of these chemical species is not captured, but this is arguably unnecessary given that we are attempting to fit just a small number (11) of spectral points.

\begin{figure}
    \centering
    \includegraphics[width=0.9\textwidth]{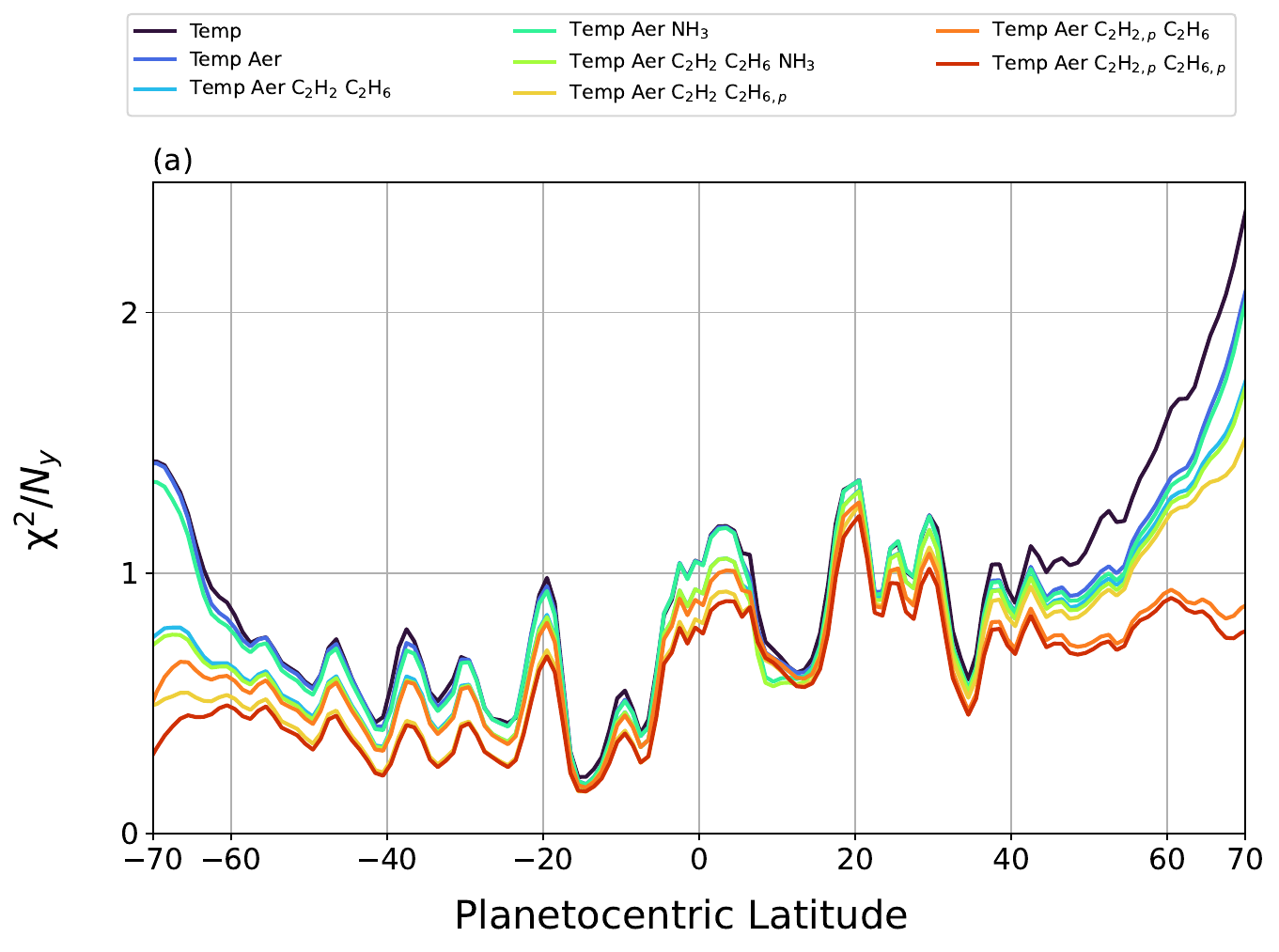}
    \caption{Latitudinal variations of the $\frac{\chi^2}{N_y}$ parameter to determine the goodness-of-fit for (a) all retrieval tests (summarised in Table \ref{tab:meridian_retrieval_tests}); and (b) a re-plot of the best fits for the two retrieval schemes that respect the $\frac{\chi^2}{N_y}\leqslant$~1 condition for the entire latitude range.}
    \label{fig:meridian_retrieval_chisquare}
\end{figure}

To determine the goodness-of-fit for each test retrieval to fit the observed meridional profiles of radiance (Figure \ref{fig:calib_profile}), the reduced chi-squared value, $\frac{\chi^2}{N_y}$, is calculated at each latitude point (Figure \ref{fig:meridian_retrieval_chisquare}). 
Here ${N_y}$ represents the number of spectral points. 
If the retrieval test (i.e., the combination of temperature, aerosol and chemical species profiles) is a good approximation to the observations, the value of the reduced chi-squared should be approximately unity $\frac{\chi^2}{N_y}$~=~1. 
However, if the fitting ``atmosphere'' (composed of the prior profiles of temperature, aerosol layer and eventually some chemical species we had allowed to be scaled or modified by the parametric setting) is not appropriate to describe the observations, the value of $\frac{\chi^2}{N_y}$ is greater than 1 because the model is incapable of fitting the data within the measurement uncertainties. 
Hence, only retrieval tests with $\frac{\chi^2}{N_y}\leqslant$~1 along the larger latitude range are considered as a good fit to the VLT/VISIR 2018 May 24$^{th}$-27$^{th}$ observations.
Both \textbf{Temp Aer C$_2$H$_{2, P}$ C$_2$H$_{6}$} and \textbf{Temp Aer C$_2$H$_{2, P}$ C$_2$H$_{6, P}$} retrievals  (i.e., both with a parametric retrieval approach of acetylene, respectively with and without a scaling of ethane profile) depict a meridional profile of $\chi^{2}/N_y$ lower or equal to 1 for almost the entire latitude range (see Figure \ref{fig:meridian_retrieval_chisquare}).  
The main benefit of using the parametric approach for acetylene and ethane occurs in polar regions, poleward 50$^{\circ}$N and 60$^{\circ}$S, where the vertical gradients of hydrocarbons are not well constrained. 
Indeed, the complex nature of auroral-related chemistry and auroral-related charged particles interaction with the neutral polar atmosphere of Jupiter, as well as the 3D extension of those processes out of the auroral oval remains poorly characterized.
The use of the simple shaped of acetylene and ethane vertical profiles through the parametric approach overcomes this lack of vertical constraints, and allow NEMESIS to move away from the prior profiles, particularly well designed from low-latitudes.

To keep the retrieval settings as simple as possible without unnecessary extra parameters, we have selected the Temp Aer C$_2$H$_{2, P}$ C$_2$H$_{6}$ test as the best one.


\section{Zonal-Mean Temperatures, Aerosols, and Composition from Pole to Pole}
\label{sec:belts_zones}
\subsection{Temperature and aerosol opacity}
 
\begin{figure}
    \centering
    \includegraphics[width=0.98\textwidth]{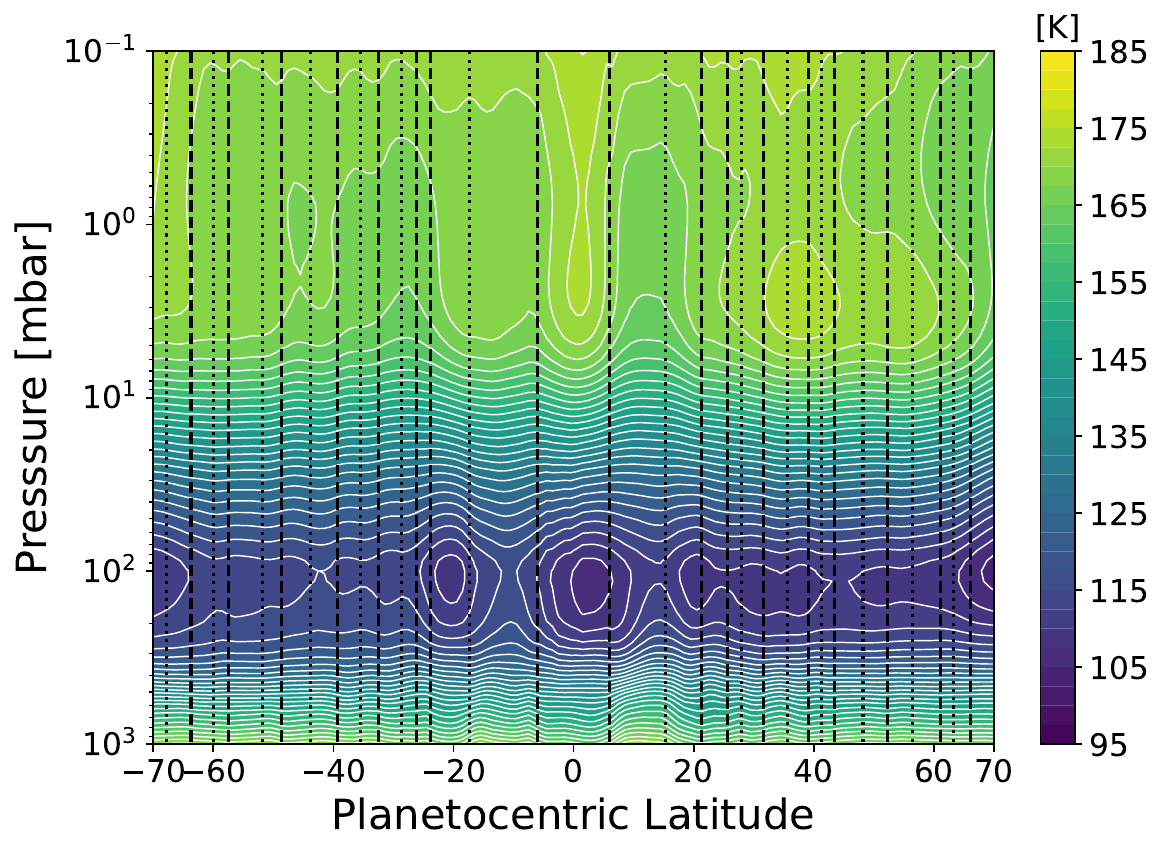}
    \caption{Pressure-Latitude cross-section of retrieved temperature from the VLT/VISIR May 2018 24$^{th}$-27$^{th}$ dataset, superposed with the location of the observed eastward zonal wind (dashed black lines) and westward zonal wind (dotted black lines) peaks \cite{Porc:03}.}
    \label{fig:zonal_temperature_cross_sec}
\end{figure}

\begin{figure}
    \centering
    \includegraphics[width=0.98\textwidth]{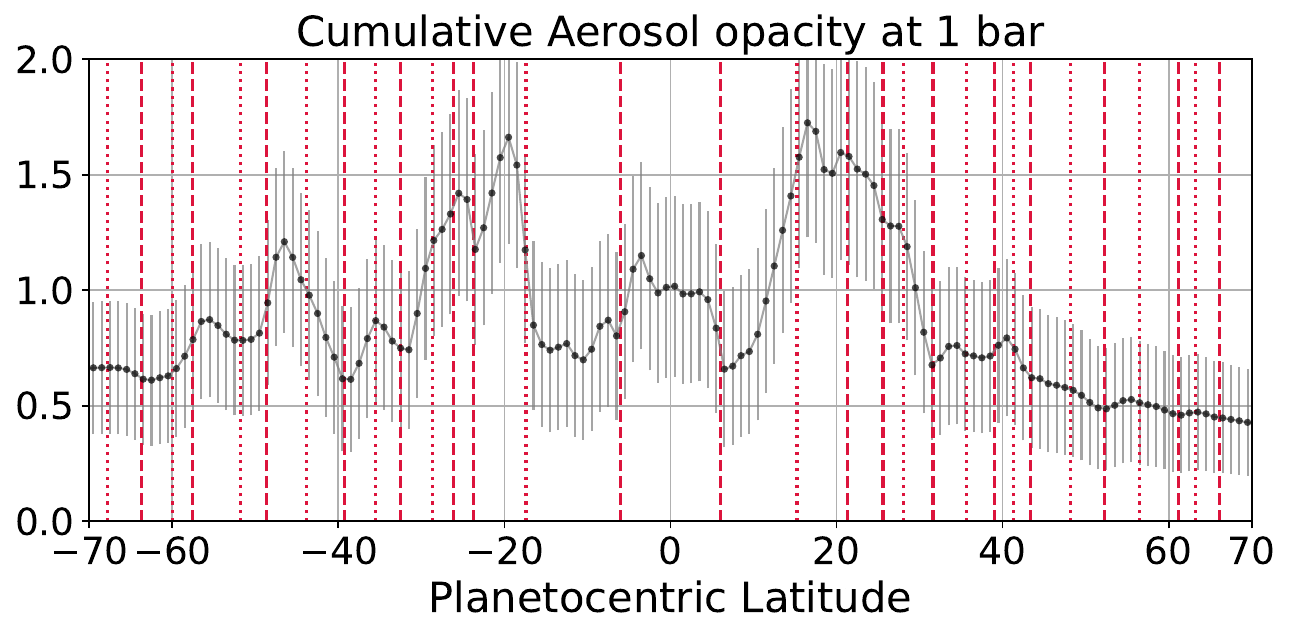}
    \caption{Cumulative aerosol opacity at 1~bar, superimposed with the location of the observed eastward zonal wind (dashed red lines) and westward zonal wind (dotted red lines) peaks \cite{Porc:03}.}
    \label{fig:aerosol_opacity}
\end{figure}

\begin{figure}
    \centering
    \includegraphics[width=0.7\textwidth]{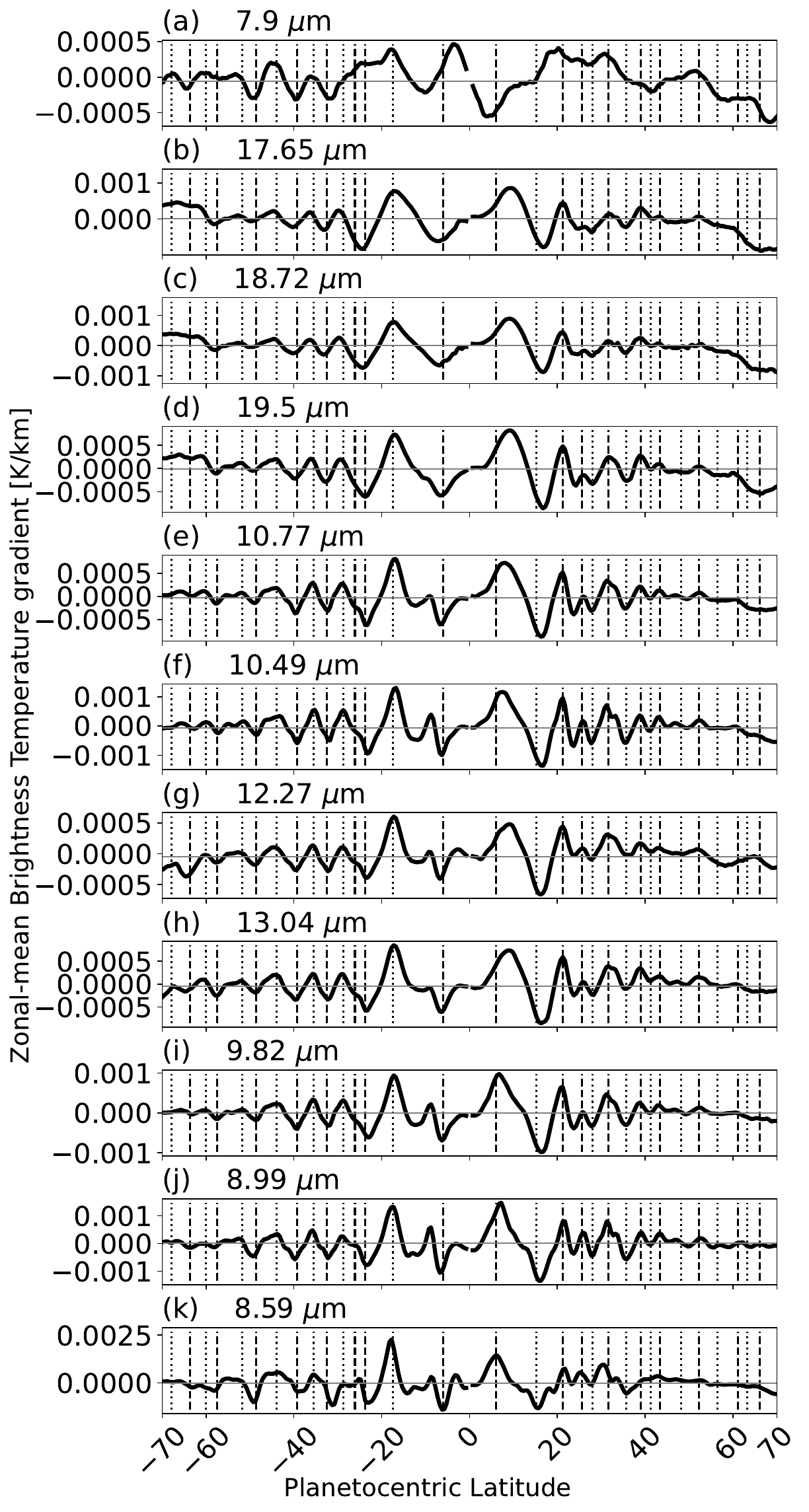}
    \caption{Meridional profiles of brightness temperature gradient calculated from the VLT/VISIR global maps for each filter, superimposed with the location of the observed eastward zonal wind (dashed lines) and westward zonal wind (dotted lines) peaks \cite{Porc:03}.}
    \label{fig:meridional_gradient_brightness_temperature}
\end{figure}

Retrieved temperatures as a function of latitude and pressure, resulting from the Temp Aer C$_2$H$_{2, P}$ C$_2$H$_{6}$ retrieval run, capture most of the upper-tropospheric and stratospheric phenomena on Jupiter (Figure \ref{fig:zonal_temperature_cross_sec}). 
For example, between 10$^{\circ}$N and 10$^{\circ}$S, we observe a cooling over the Equatorial Zone, directly surrounded by the warm North Equatorial Belt and the South Equatorial Belt (NEB and SEB, $\sim$7 to 17-19$^{\circ}$ of latitude), and the cold North Tropical Zone and the South Tropical Zone (from 17 to 24$^{\circ}$N and from 19 to 26$^{\circ}$S of latitude).
The high spatial resolution of the VISIR maps reveals that this pattern of cool zones and warm belts continues all the way to the boundary of polar domains (i.e., up to $\pm$60$^{\circ}$).
In addition to the thermal contrast of the belt/zone signature, the VLT/VISIR dataset also captures the aerosol opacity variations, with zones possessing a higher aerosol opacity compared to their neighbouring belts, suggesting condensation of saturated vapours in zones and sublimation/evaporation in cloud-free belts (Figure \ref{fig:aerosol_opacity}).  
Somewhat surprisingly, the equatorial aerosol opacity is not the largest on the planet, as was observed in previous studies \cite<e.g.,>[]{Flet:16jupiter} - it is possible that the effects of the Equatorial Disturbance with visibly orange cloud cover in 2018-19 \cite{Antu:20} had reduced the equatorial opacity to lower levels than normal.
The transition locations between cold-cloudy zones and warm-cloud-free belts are not clearly seen in this retrieved temperature cross-section, as the retrieval calculations smooth some of the observed features. 
Nonetheless, by calculating the meridional thermal gradients from the limb-corrected brightness temperature observations for each filter (i.e., from the zonal-mean meridional profiles of brightness temperature calculated from the global maps shown on Figure \ref{fig:global_maps}), we can easily assess whether the observed thermal field follows the belts and zones all the way up to the highest latitudes.
Figure \ref{fig:meridional_gradient_brightness_temperature} shows a correlation with the locations of the thermal gradient and zonal jet peaks in the troposphere, particularly clearer for the southern hemisphere (the one facing the Earth when this ground-based observations dataset was acquired). 
In the troposphere, for 17.65 to 8.59-$\mu$m filters (Figure \ref{fig:meridional_gradient_brightness_temperature}b-\ref{fig:meridional_gradient_brightness_temperature}k) zones have eastward-propagating jets (prograde, their peaks are depicted by dashed lines on Figures \ref{fig:zonal_temperature_cross_sec} and \ref{fig:aerosol_opacity}) at their polar edges whereas belts exhibit westward-propagating jets (retrograde, their peaks are depicted by dotted lines on Figures \ref{fig:zonal_temperature_cross_sec} and \ref{fig:aerosol_opacity}) along their polar boundaries.

Moreover, the retrieved temperature cross-section (Figure \ref{fig:zonal_temperature_cross_sec}) shows clear cooling poleward of 60$^{\circ}$N and 60$^{\circ}$S, between 300 and 30~mbar, characteristic of the cold polar vortices in each hemisphere.  
This is driven primarily by the Q-band observations, and the clear boundary at high latitudes in Figure \ref{fig:global_maps} which is likely to be associated with the highest-latitude jets at $68^\circ$. 
To better emphasise these polar vortex boundaries, we refer to the meridional gradient of brightness temperature for the 17.65-$\mu$m filter (Figure \ref{fig:meridional_gradient_brightness_temperature}b), wavelength for which the polar vortices is best defined on the global maps. 
In the southern hemisphere, the poleward cooling occurs at the equatorward edge of the southernmost westward jet, i.e., at 67$^{\circ}$S, up to the pole.
For the northern hemisphere, the location of the polar vortices is less evident. 
For the 7.9-$\mu$m filter, the northernmost poleward cooling occurs at the northernmost westward jet location (64$^{\circ}$N), whereas the 17.65-$\mu$m filter depicts this cooling at second northernmost eastward jet location (62$^{\circ}$N).
The aerosol distributions in the polar domains are somewhat asymmetric in cloud coverage (Figure \ref{fig:aerosol_opacity}).
There is a net decrease of the northern aerosol opacity, about 0.5 in the north, whereas the southern pole retains an aerosol opacity of about 0.6, implying increased aerosol opacity over the southern pole compared to the northern one, though the north and south aerosol optical depths are equal within the retrieval uncertainties.

\begin{figure}
    \centering
    \includegraphics[width=0.8\textwidth]{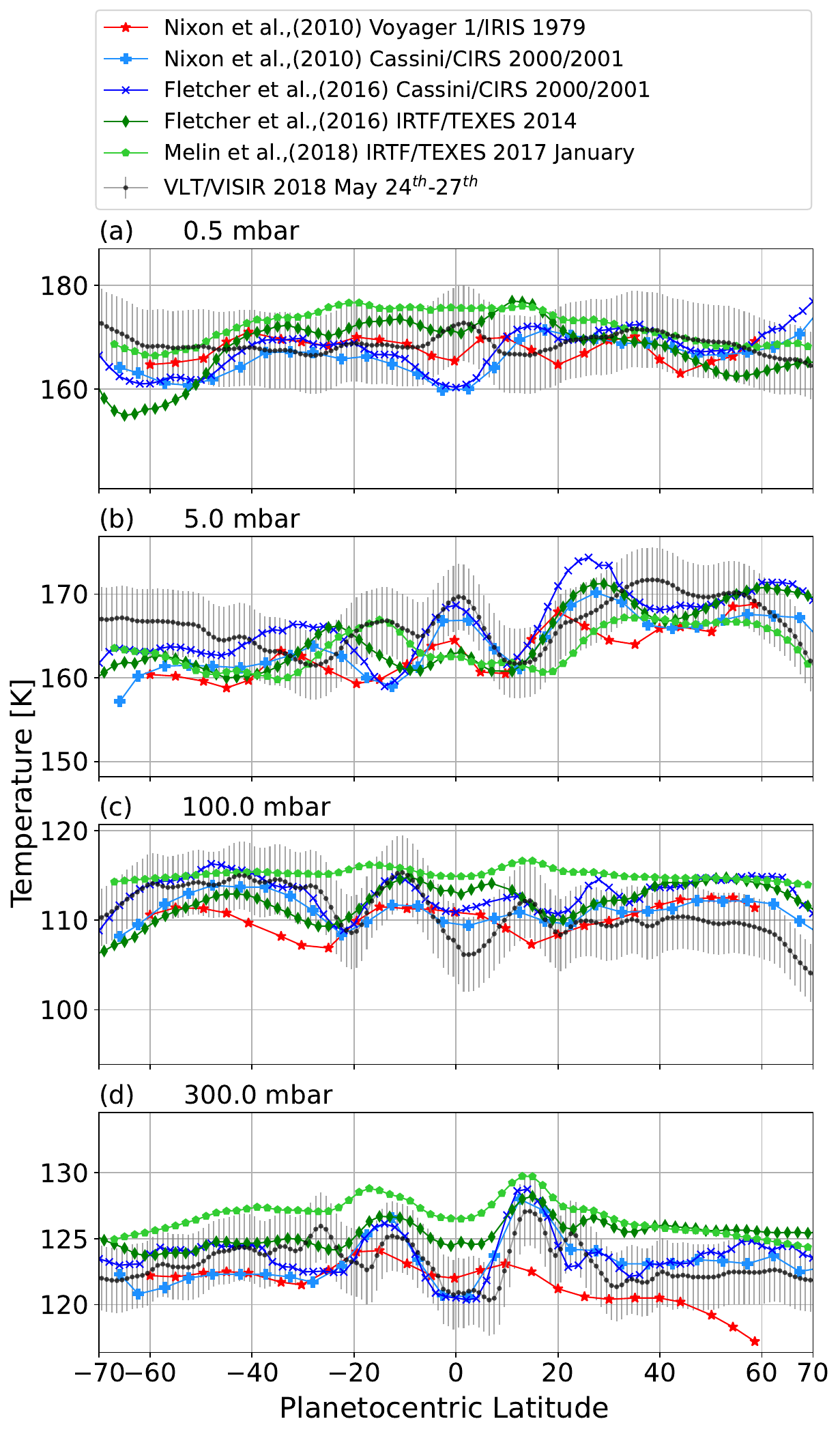}
    \caption{Zonal-mean retrieved temperatures at four pressure levels spanning the troposphere and stratosphere from the VLT/VISIR May 2018 24$^{th}$-27$^{th}$ dataset, compared with observations from Voyager/IRIS, Cassini/CIRS, and IRTF/TEXES.}
    \label{fig:temperature_merid_profiles}
\end{figure}

Comparison with flyby observations of the Voyager-1/IRIS and Cassini/CIRS spectrometers, as well as ground-based observations by the IRTF/TEXES instrument (Figure \ref{fig:temperature_merid_profiles}), shows an overall consistency between our retrieval results and previous studies.
In the equatorial stratosphere, the retrievals reveal a warm temperature anomaly in the 0.5-5.0 mbar range centered at the equator, surrounded by two cooler anomalies centered at $\pm$15$^{\circ}$ latitude (Figure \ref{fig:temperature_merid_profiles}a). 
Conversely, flyby observations from Voyager-1/IRIS in 1979 and Cassini/CIRS in 2000/2001, as well as IRTF/TEXES measurements in 2014 \cite{Nixo:10,Flet:16jupiter,Meli:18}, depict a cold equatorial region at the 0.5-mbar pressure level, and a relatively uniform meridional contrast of the equatorial stratosphere in the Gemini/TEXES measurements of March 2017 \cite{Flet:20Jupiter_equatorialplume}.  
The transition from cold to warm equatorial stratospheric temperatures between 2014 and 2018, along with the off-equatorial thermal anomalies near $\pm15^\circ$, are the manifestations of Jupiter's equatorial ``Quasi-Quadrennial Oscillation'' \cite{Leov:91,Orto:91,Frie:99,Simo:07,Cose:17,Cose:20,Antu:21}. 
An asymmetric temperature changing is also present in the sub-equatorial regions (from 8 to 20$^{\circ}$), transitioning from warm to cold anomalies of temperature, with a much clearer and stronger change in the northern sub-equatorial region. 
The downward progression of the equatorial stratospheric oscillation requires more than just the May 2018 snapshot, and will be the topic of forthcoming studies.

The stratospheric temperature in the southern hemisphere does not capture the usual meridional variations at the 0.5-mbar and 5-mbar pressure levels shown in Figures \ref{fig:temperature_merid_profiles}a and \ref{fig:temperature_merid_profiles}b. 
At 0.5~mbar, the retrieved temperature from VLT/VISIR measurements varies from 20$^{\circ}$S to 65$^{\circ}$S with an almost constant gradient, without showing the large decrease of about 10-20~K from 35$^{\circ}$S to 60$^{\circ}$S as shown in Voyager-1/IRIS, Cassini/CIRS and IRTF/TEXES profiles.    

At 5~mbar, the temperature of the southern hemisphere from 40$^{\circ}$S to 70$^{\circ}$S is around 5~K warmer than previous studies, with a flat variation over latitude. 
Moreover, the stratospheric northern mid-latitudes, between 35 to 60$^{\circ}$N, show local warm regions in the VLT/VISIR profiles at 5~mbar, centered at 40 and 55$^{\circ}$N (Figure \ref{fig:temperature_merid_profiles}b).
Those warm regions are not consistent with previous observations, but this could well be the signature of stratospheric mid-latitude variability associated with the propagation of planetary-scale waves (as a Rossby planetary-scale wave, mentioned in section \ref{sec:obs_processing_method} and shown on Figure \ref{fig:Rossby_waves_patterns_TB_perturbation}).

From 70$^{\circ}$N to 70$^{\circ}$S, VLT/VISIR temperatures have the same variations in the troposphere as previous observations from Voyager-1/IRIS, and especially from Cassini/CIRS and IRTF/TEXES (Figures \ref{fig:temperature_merid_profiles}c and \ref{fig:temperature_merid_profiles}d).
Both 100-mbar and 300-mbar pressure levels reveal a cold Equatorial Zone surrounded by the warm North and South Equatorial Belts, producing a 7~K-amplitude variation in the meridional profile of temperature from 20$^{\circ}$N to 20$^{\circ}$S.
Temperatures consistently follow the Cassini/CIRS observations at 300~mbar, but its variability from 20$^{\circ}$N to 20$^{\circ}$S is amplified in the VLT/VISIR observations at 100~mbar compared to those from IRTF and spacecraft facilities, most likely a consequence of the higher spatial resolution of our observations.
Another difference occurring at low latitudes in the VLT/VISIR observations is the local maximum of temperature at 25$^{\circ}$S, producing an increased temperature of 3-4~K, which is not seen in the other datasets.

\subsection{Zonal wind shear}
\begin{figure}
    \centering
    \includegraphics[width=\textwidth]{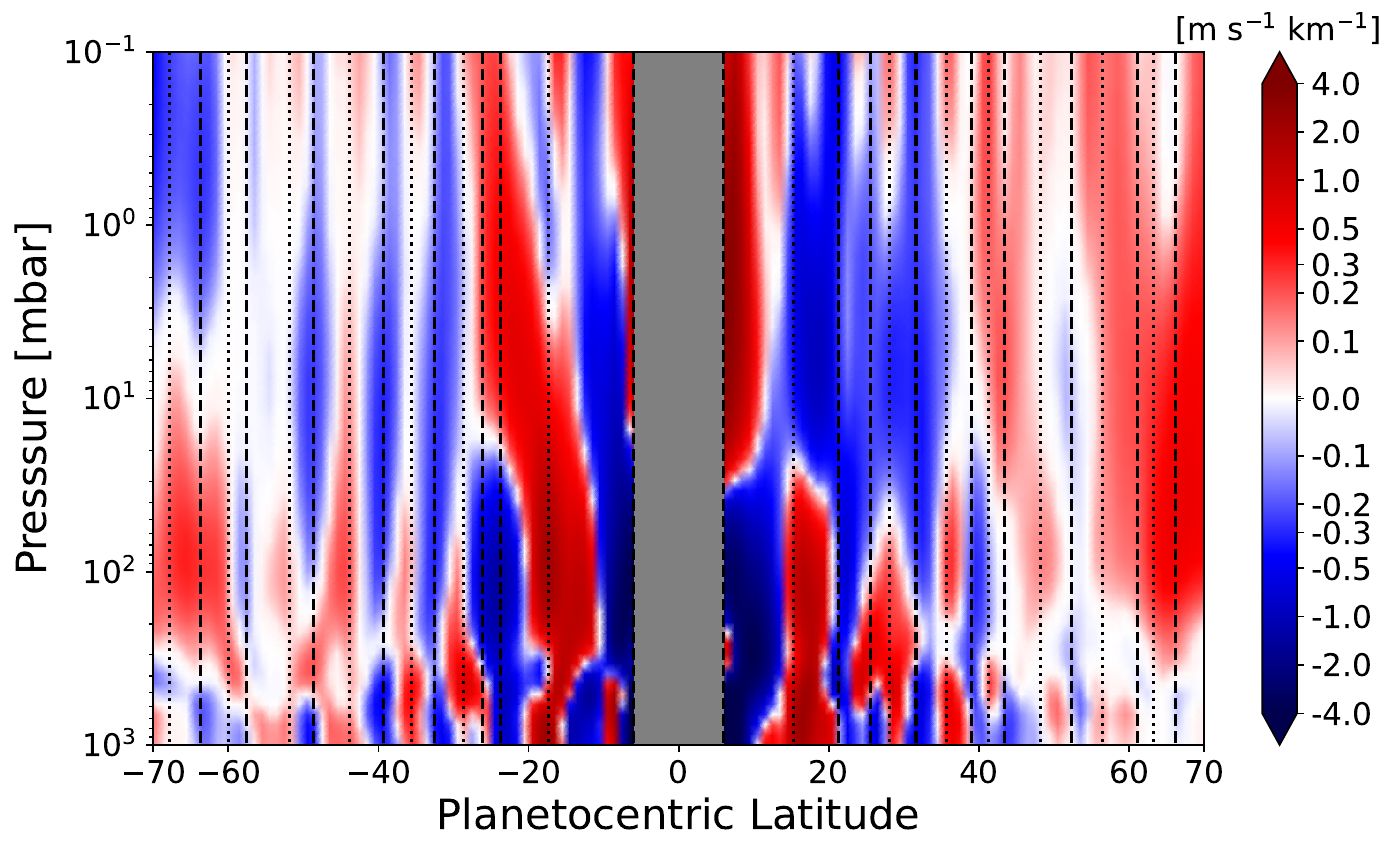}
    \caption{2D cross-section of zonal-mean zonal wind shear calculated from the VLT/VISIR retrieved temperature using the equation \ref{eq:windshear_from_temperature}, superposed with the location of the observed eastward zonal wind (dashed black lines) and westward zonal wind (dotted black lines) peaks \cite{Porc:03}. The colour scale is a symmetric logarithmic normalisation to emphasise gradients at mid and high latitudes. Tropical regions equatorward of 6$^{\circ}$ are omitted.}
    \label{fig:windshear_cross_sec}
\end{figure}

The retrieved zonal-mean temperatures and their meridional gradients can be used to estimate the vertical shear on the zonal winds, using thermal wind equation (equation \ref{eq:windshear_from_temperature}): 
\begin{equation}
    \frac{\partial \overline{u}}{\partial z} = - \frac{g}{f \overline{T}}\times\frac{\partial \overline{T}}{\partial y}
    \label{eq:windshear_from_temperature}
\end{equation}
\noindent where $\overline{u}$ is the zonal-mean zonal wind, $g$ is the gravitational acceleration, $f$ is Coriolis parameter, $\overline{T}$ is the zonal-mean temperature and $y$ the north-south distance in kilometres (Figure \ref{fig:windshear_cross_sec}). 
This approximation is not applicable to equatorial region, as the Coriolis parameter $f$ approaches zero. 
Therefore, we have omitted tropical regions equatorward of 6$^{\circ}$ of latitude.

In the lower to middle stratosphere, there is clear correlation between the location of the zonal jets defined by the dotted and dashed lines.
This correlation results in zones on the equatorward sides of the eastward jet and belts on their poleward sides; and zones on the poleward side of the westward jets, with belts on their equatorward side, between 500~mbar to the tropopause ($\sim$100~mbar). 
The correlation also appears to be somewhat stronger in the southern hemisphere than in the north, confirming the pattern highlighted in previous studies using Juno Microwave Radiometer data \cite{Flet:21}.

\subsection{Meridional distribution of hydrocarbons}
\begin{figure}
    \centering
    \includegraphics[width=\textwidth]{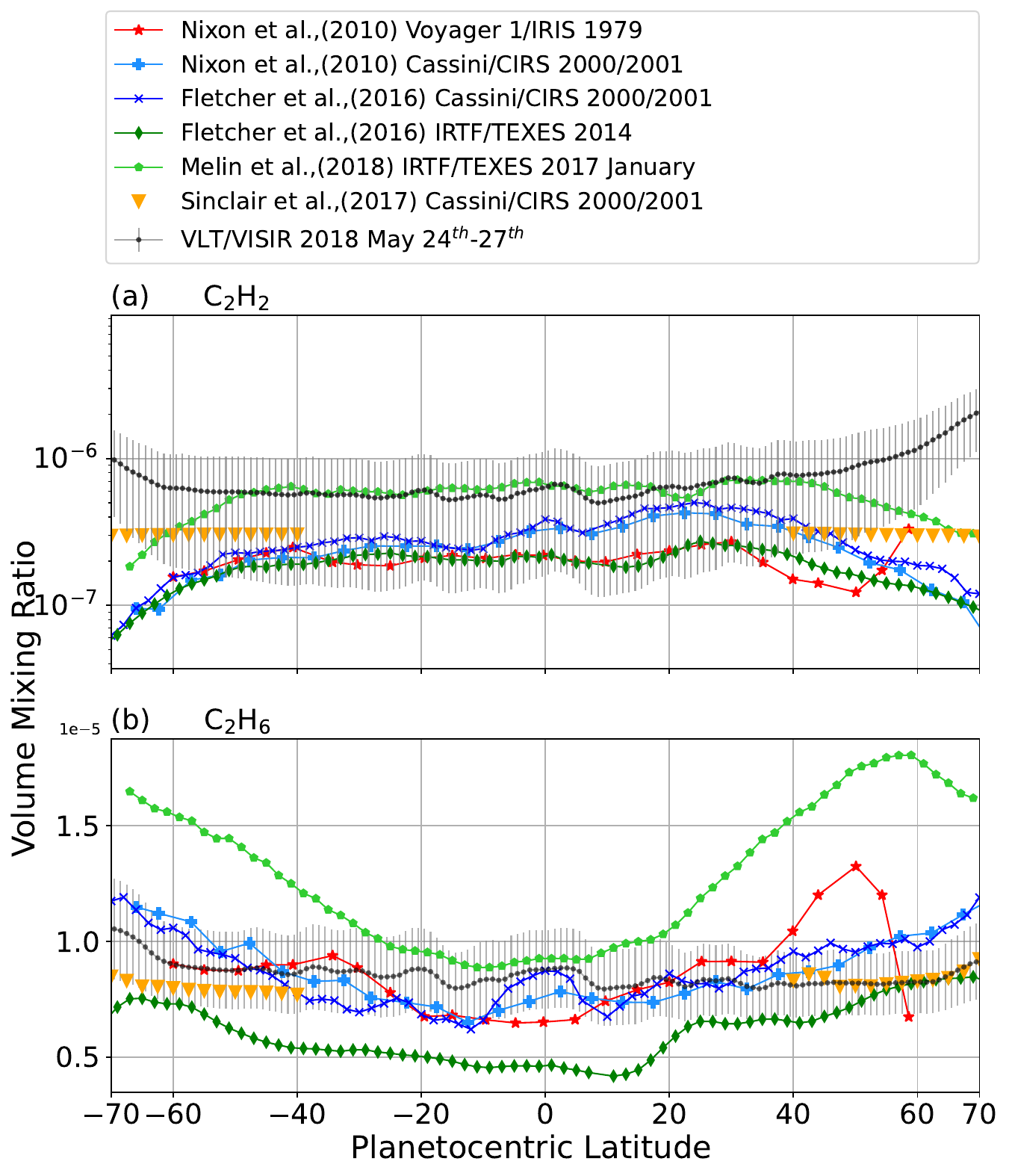}
    \caption{Retrieved meridional profiles of hydrocarbons at 1~mbar from the VLT/VISIR May 2018 24$^{th}$-27$^{th}$ dataset for the 
    \textbf{Temp Aer C$_2$H$_{2, P}$ C$_2$H$_{6}$} (black) retrieval runs.}
    \label{fig:hydrocarbons_merid_profiles}
\end{figure}

Retrieved hydrocarbons as a function of latitude, resulting from the Temp Aer C$_2$H$_{2, P}$ C$_2$H$_{6}$ retrieval run, capture most of the meridional variatbility of previous work (Figure \ref{fig:hydrocarbons_merid_profiles}). 
Compared to those studies, the VLT/VISIR observations are within the range of previous observations from several facilities, and is particularly consistent with the IRTF/TEXES 2017 dataset from \citeA{Meli:18}. 
Between 60$^{\circ}$S and 40$^{\circ}$N, acetylene distribution retrieved from the VLT/VISIR dataset describes an almost flat gradient, with a variability not larger than 2$\times$10$^{6}$. 
\begin{figure}
    \centering
    \includegraphics[width=\textwidth]{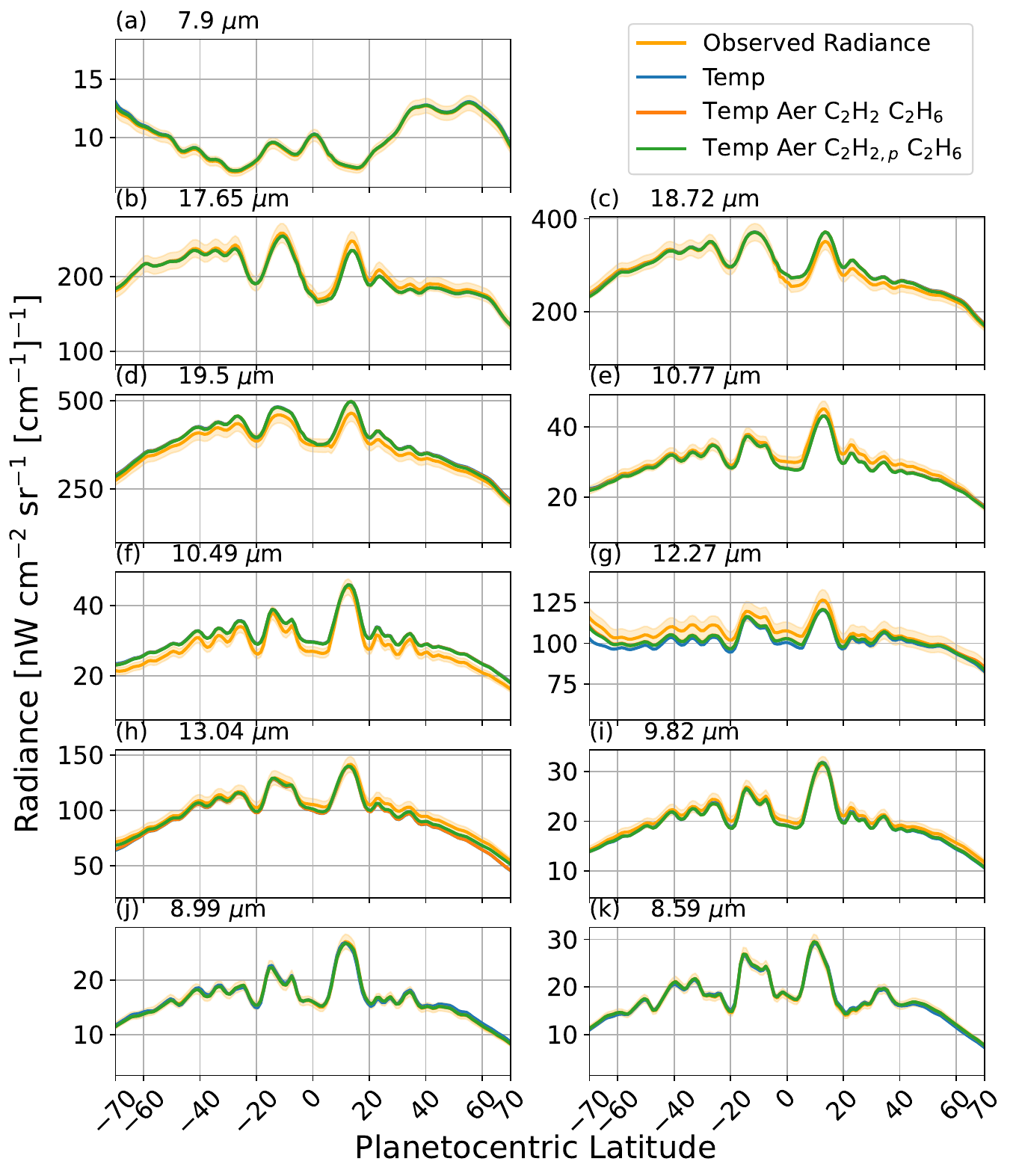}
    \caption{Comparison of observed (orange) and retrieved meridional profiles of radiance for three case tests: \textbf{Temp} (in blue), \textbf{Temp Aer C$_2$H$_{2}$ C$_2$H$_{6}$} (in dark orange), and \textbf{Temp Aer C$_2$H$_{2, P}$ C$_2$H$_{6}$} (in green) retrieval runs to highlight the need for the parametric retrieval approach for hydrocarbons to better fit the southern auroral-related increase in the radiance profiles.}
    \label{fig:obs-retrieved_radiance_comparison_merid_profiles}
\end{figure}
To correctly fit the VLT/VISIR observations at all latitudes with a $\chi^{2}/N_y$ less than 1, we have to adopt the C$_2$H$_2$ parametric and C$_2$H$_6$ scaling retrieval technique, particularly for the high-latitude regions (poleward of 60$^{\circ}$).
However, the best fits to the data imply polar enhancements of almost an order of magnitude for C$_2$H$_2$, which has not been seen in previous retrieval analyses to date (Figure \ref{fig:hydrocarbons_merid_profiles}). 
This is needed because the temperature-only fit (Figure \ref{fig:obs-retrieved_radiance_comparison_merid_profiles}, blue line, based largely on the 7.9-$\mu$m brightness) produces a stratospheric temperature that is too cool to explain the observed emission in the ethane- and acetylene-sensing filters (i.e. 12.27- and 13.04-$\mu$m filters), so the abundances have to increase significantly to compensate.

However, the large enhancement of volume mixing ratio in high-to-polar latitudes is not found in the meridional distribution of ethane, where a scaling retrieval settings was enough to fit the observations (i.e., with a $\chi^{2}/N_y$ less than 1 for the whole latitude range). 
Global meridional variations of ethane, from 70$^{\circ}$S to 70$^{\circ}$N exhibit a slight increase of volume mixing ratio toward the higher latitudes. 
This is consistent with Voyager and Cassini previous observations, albeit quite reduced. This is particularly true for the northernmost latitudes, producing an asymmetry in the high latitude concentration of ethane. 
This asymmetry might simply result from the observational conditions, i.e. the sub-observer point at -3.76$^{\circ}$ offered a better view of the southern pole.

In the high latitude to polar region, the vertical structure of hydrocarbons is still rather poorly known, as well as their interactions with the auroral-related particle precipitation, their photochemical transformations, their transport across the polar vortex, etc. 
It is possible that our form of parameterisation is simply inappropriate for the polar domain.
Although the bright polar emission at 13 $\mu$m (Figure \ref{fig:calib_profile}h) is most readily explained by a significant increase in acetylene abundance at high latitudes, this is not supported by previous measurements, and points to the extreme challenge of constraining polar temperatures and hydrocarbons using filtered imaging alone.

\section{Structure of the Great Red Spot in 2018}
\label{sec:GRS}

\begin{figure}
    \centering
    \includegraphics[width=\textwidth]{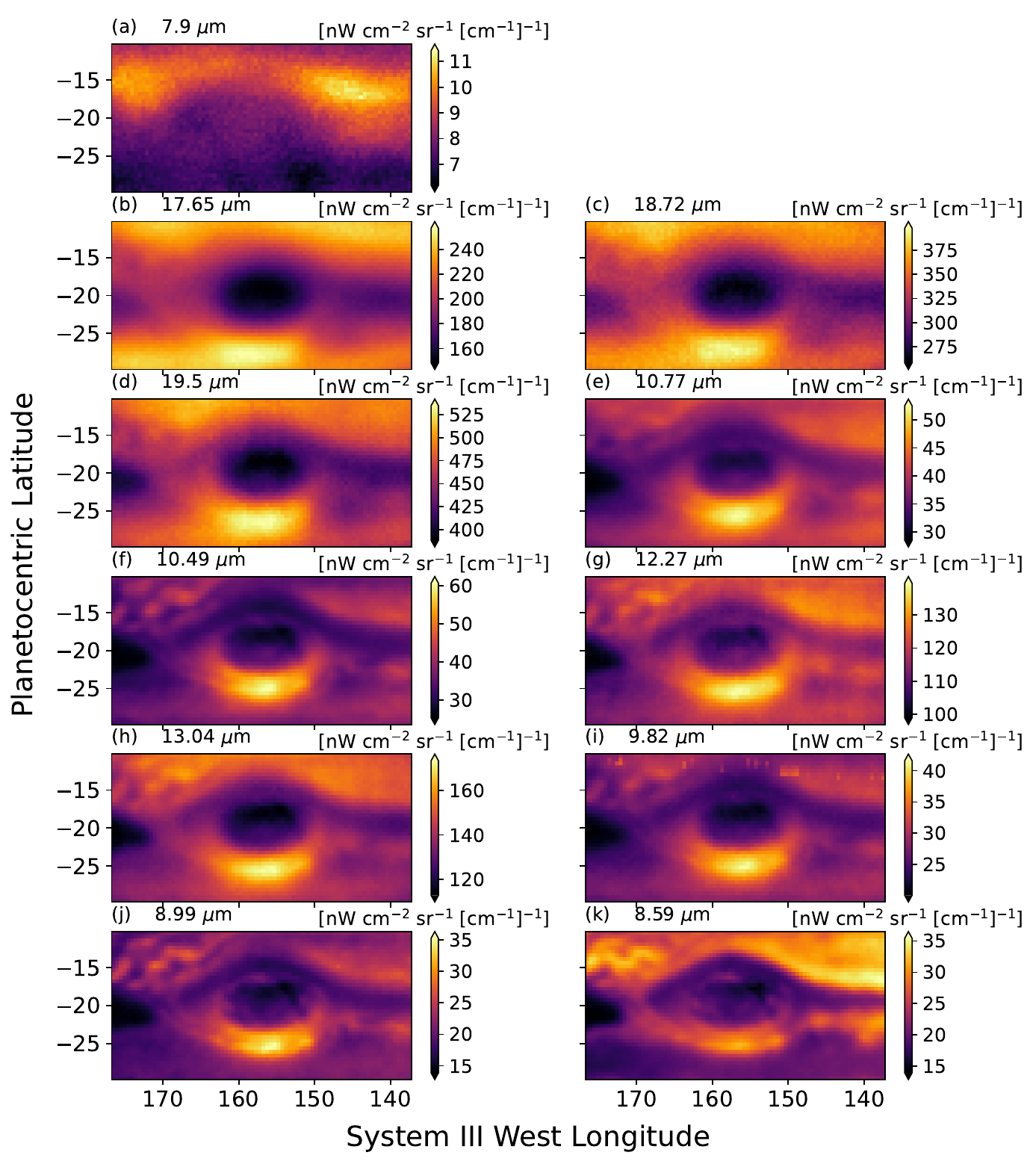}
    \caption{Great Red Spot radiance maps for VLT/VISIR (2018 May 24$^{th}$-27$^{th}$) N-Band (7.9-13.04 $\mu$m) and Q-Band (19.50, 18.72 and 17.65 $\mu$m) observations.}
    \label{fig:GRS_radiance}
\end{figure}

The VLT/VISIR 2018 May 24$^{th}$-27$^{th}$ dataset offers excellent coverage of the Great Red Spot (GRS), in particular on the night of May 26$^{th}$, and it provides an opportunity to study its thermal and cloud structures. 
Thanks to the diffraction-limited filtered imaging at high sensitivity for a wavelength range of 4.82--19.5$\mu$m of the VISIR plate-scale (0.045''/pixel), and using the Data Reduction Manager pipeline, we have interpolated the radiance maps onto a 0.5-degree grid, defined as the same across all filters, which oversamples the longest wavelengths. 
Hence, from a single set of images that captures the best view of the GRS during the night of 2018 May 26$^{th}$, we have selected an area of 20$^{\circ}$-latitude by 40$^{\circ}$-longitude centered at latitude 20$^{\circ}$S -- longitude 157$^{\circ}$W System III (coordinates of GRS center in May 2018), shown on a $0.5\times0.5^\circ$ grid in Figure \ref{fig:GRS_radiance}. 
We have made a retrieval calculation for each single pixel in this area, permitting a longitude-latitude resolved retrieval of the GRS.

\subsection{Interior structure of the Great Red Spot}

\begin{figure}
    \centering
    \includegraphics[width=\textwidth]{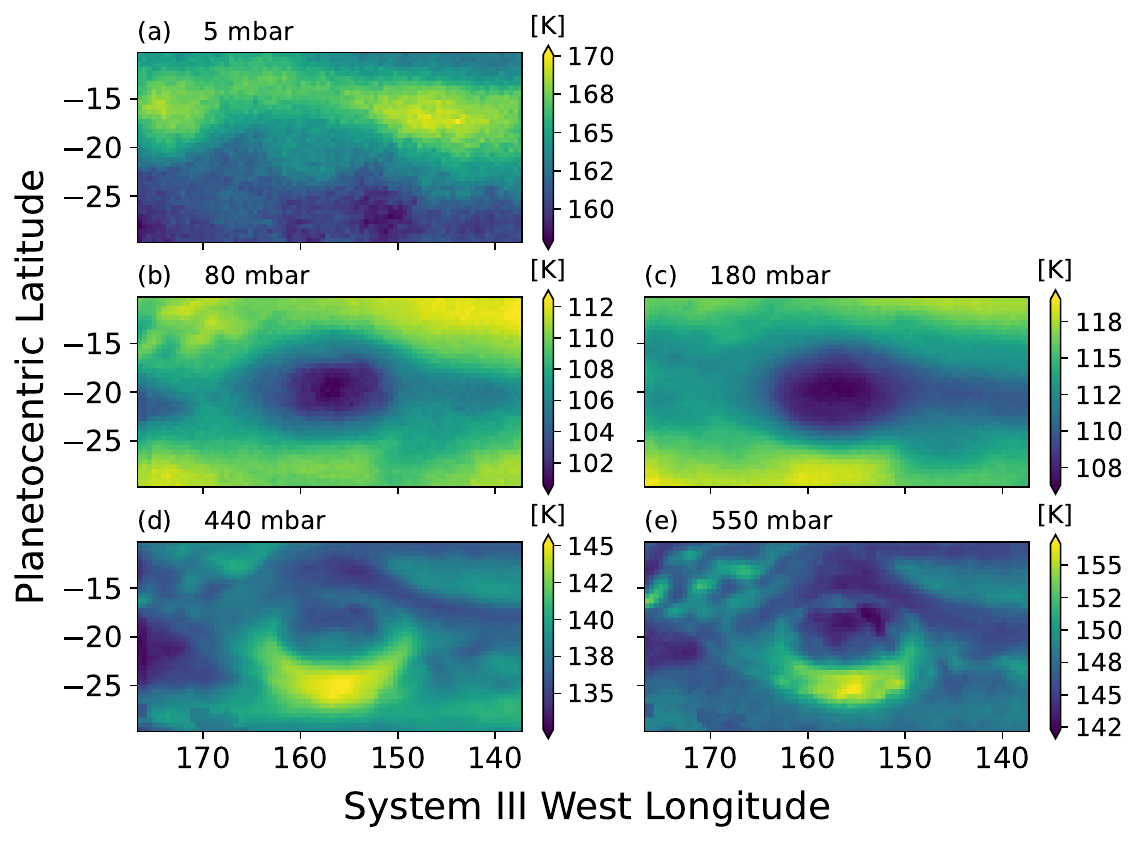}
    \caption{Great Red Spot retrieved temperature maps from VLT/VISIR 2018 May 26$^{th}$ dataset.}
    \label{fig:GRS_temperature}
\end{figure}

\begin{figure}
    \centering
    \includegraphics[width=\textwidth]{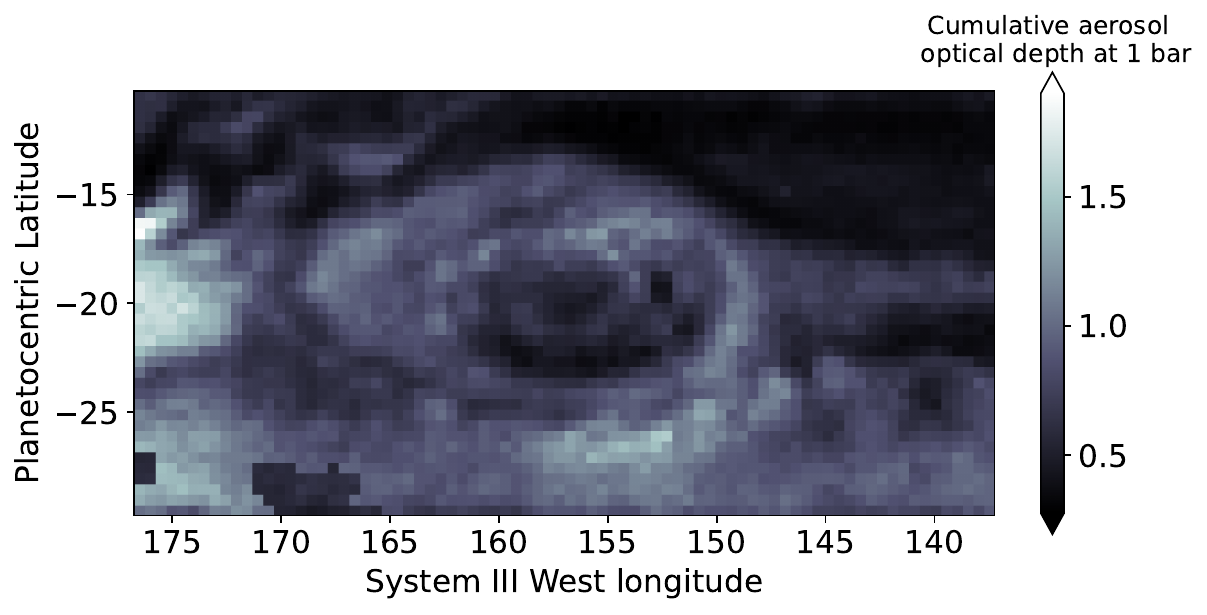}
    \caption{Great Red Spot map of the cumulative aerosol optical depth at 1~mbar from VLT/VISIR 2018 May 26$^{th}$ dataset retrieval calculations.}
    \label{fig:GRS_aerosols}
\end{figure}

Retrieved temperatures over the Great Red Spot (Figure \ref{fig:GRS_temperature}) from the VLT/VISIR data shows that the GRS is a cold anticyclonic feature, but with the continued existence of a tropospheric ``warm spot'' on the inner edge of the core at 21$^{\circ}$S, 153$^{\circ}$W System III, suggesting that the warm spot might be long-lived but varies over short timescales, potentially related to the visibility of the red core observed in the amateur observation images in Figure \ref{fig:global_map_amateur}. 
The location of this ``warm spot'' is possibly related with the location of \textbf{where} the tangential winds of the GRS stagnate and even reverse to become cyclonic \cite{Flet:10grs}. 
Also evident in the radiance maps (Figures \ref{fig:GRS_radiance}f to \ref{fig:GRS_radiance}k), the localised ``warm spot'' is consistent with previous retrievals from mid-IR imaging data \cite{Flet:10grs}.
However, the ``warm spot'' in the core is not seen at the tropopause levels, from the 180-mbar to the 80-mbar pressure levels, where the contrast shows predominantly a cool core with a warm periphery (Figures \ref{fig:GRS_temperature}b-\ref{fig:GRS_temperature}c); nor in the stratospheric 5-mbar pressure level, where the GRS shape is barely distinguishable (Figure \ref{fig:GRS_temperature}a). 
From 380 to 600~mbar, this ``warm spot'' appears to drift eastward, from 157 to 153$^{\circ}$W System III with increasing pressure (Figures \ref{fig:GRS_temperature}d-\ref{fig:GRS_temperature}k), behavior also noticed in radiance maps (Figures \ref{fig:GRS_radiance}e-\ref{fig:GRS_radiance}j).
The largest contrast of the ``warm spot'' occurs at 550~mbar, with a maximum value of $\sim$150~K and a contrast of 5-7~K warmer over the inner coldest temperatures. 
At this location, the aerosol distribution shows a depletion of aerosols (Figure \ref{fig:GRS_aerosols}). 
The localised warming sensed at the cloud-top level (Figure \ref{fig:GRS_radiance}k) and obtained in the retrieved temperature maps may result in evaporation of any condensed ices there, implying a relatively cloud-free region on the southern inner edge of the GRS, at least at the 500-700 mbar level.

The core of the GRS at 550~mbar represents the coldest part of this region by $\sim$10~K, reflecting the intense upwelling that is characteristic of the GRS. 
Regarding the aerosol distribution resulting from the retrieval calculations (Figure \ref{fig:GRS_aerosols}) shows that the coldest parts of the GRS core are correlated with some of the highest aerosol opacities in the southern tropics. 
This is likely due to enhanced condensation of aerosols at low temperatures, or due to the production of photochemical aerosols (i.e., the red chromophore) entrained by the peripheral jet of the GRS, confirmed by the visible amateur observations showing three different regions of reddish color (Figure \ref{fig:global_map_amateur}).
In visible light, the core of the GRS is composed of a central area of dark red chromophores, surrounded by a dark orange area, which is finally encircled by dark red chromophores, with the close match to the VISIR aerosol map suggesting a clear distinction between these three regions based on temperature (i.e., three different regions of condensation, and/or production of photochemical aerosols).

The warm southern periphery varies from 145~K at 440 mbar to 157~K at 550~mbar, up to $\sim$5~K warmer than the surrounding Southern Tropical Zone. 
This region of increased warming on the southern edge of the GRS is one of the warmest features in the region, beginning within the annulus at the 380-mbar level and extending beyond the southern rim of the vortex to the cloud-top level (Figures \ref{fig:GRS_temperature}d-\ref{fig:GRS_temperature}k).
As with the tropospheric aerosol opacity obtained with VISIR in 2006--2008 observations \cite{Flet:10grs}, the VISIR 2018 observations reveal a well-defined lane of depleted aerosols around the GRS periphery (Figure \ref{fig:GRS_aerosols}). 
The aerosol lane depletion also corresponds to a thin line of cooling in temperature maps, splitting the warm southern periphery around the 500-mbar level in two warm peripheries and afterward acting as a barrier between the actual GRS periphery and this thin warm line wrapping around the GRS for the deepest pressure levels.

\subsection{Great Red Spot environs}

The complex interaction between the eastward and westward jets bordering the SEB, and the anticyclonic annulus of the GRS, leads to turbulent rifting northwest of the GRS in a region called the GRS wake. 
At this time, visible-light observations also showed an unusual dark feature (``South Tropical Disturbance'') on the W side of the GRS, through which turbulence was streaming southward then eastward around the GRS (Figure \ref{fig:global_map_amateur}). 
This appears as a warm extension of the SEB in Figures \ref{fig:GRS_radiance} and \ref{fig:GRS_temperature}.
The wake region displays a range of temperature contrasts related to small-scale plumes in the local region of 0.5-1 K at 440 mbar, but higher at 80 mbar the region exhibits little spatial contrast in temperature.
The regions of greatest aerosol depletion are in the South Equatorial Belt (SEB) directly to the east and west of the GRS and co-located with the warmest temperatures (South Equatorial Belts lane) and strong vertical mixing (GRS wake). 
The perennial turbulence in the GRS wake creates chaotic structures of upward and downward motion (cooling and warming air) resulting in changes in aerosol opacity. 
This activity in the South Equatorial Belt is observed from the 10.77 to 8.59-$\mu$m images (Figures \ref{fig:GRS_radiance}e-\ref{fig:GRS_radiance}k, respectively) as strong variations in flux, implying strong aerosol clearing and production.

The thermal field at 5 mbar (Figure \ref{fig:GRS_temperature}a) displays regions of elevated temperature in the SEB immediately north-east and north-west of the GRS, with peak temperatures at 15 and 17$^{\circ}$S, each $\sim$10$^{\circ}$-wide in longitude and estimated to be at 170~K, $\sim$4 K warmer than the surroundings.  
They lie on both sides of the northern periphery of the GRS, with the western warm spot directly on top of the GRS wake.  
It is possible that this warm stratospheric band is being influenced by the anticyclone in the troposphere below.   
Thermal waves near the tropopause and in the lower stratosphere can be generated by tropospheric meteorology and convective plumes \cite{Andr:87} and disruption of flows around vortices \cite{Li:06}. 
It is plausible that these stratospheric hot spots are the result of intense convection and turbulent mixing in the troposphere that generates atmospheric waves that propagate vertically and break at some critical altitude depositing their kinetic energy into these higher altitudes as thermal energy, warming the region. 
The stratospheric temperatures above the GRS have not been previously studied \cite{Flet:10grs} and would benefit from a systematic study over multiple epochs to see if this structure of warm stratospheric perturbations is a persistent feature.

\section{Polar dynamics}
\label{sec:polar_regions}
  
Observing the polar regions of Jupiter via Earth-based facilities is a significant challenge, as they are viewed at an oblique angle with a large emission angle, inducing strong limb brightening or darkening at the planetary limbs. 
To highlight the polar atmospheric features, a limb-darkening correction (described in section \ref{subsec:limb_correction}) has been applied to the global maps.
In this section we characterise the polar domain of Jupiter and compare to Juno/JunoCam observations of the polar hazes (``polar hoods'') acquired during the same period.

\subsection{Jupiter's Cold Polar Vortices}
\label{polar_zonal_orga}
\begin{figure}
    \centering
    \includegraphics[width=\textwidth]{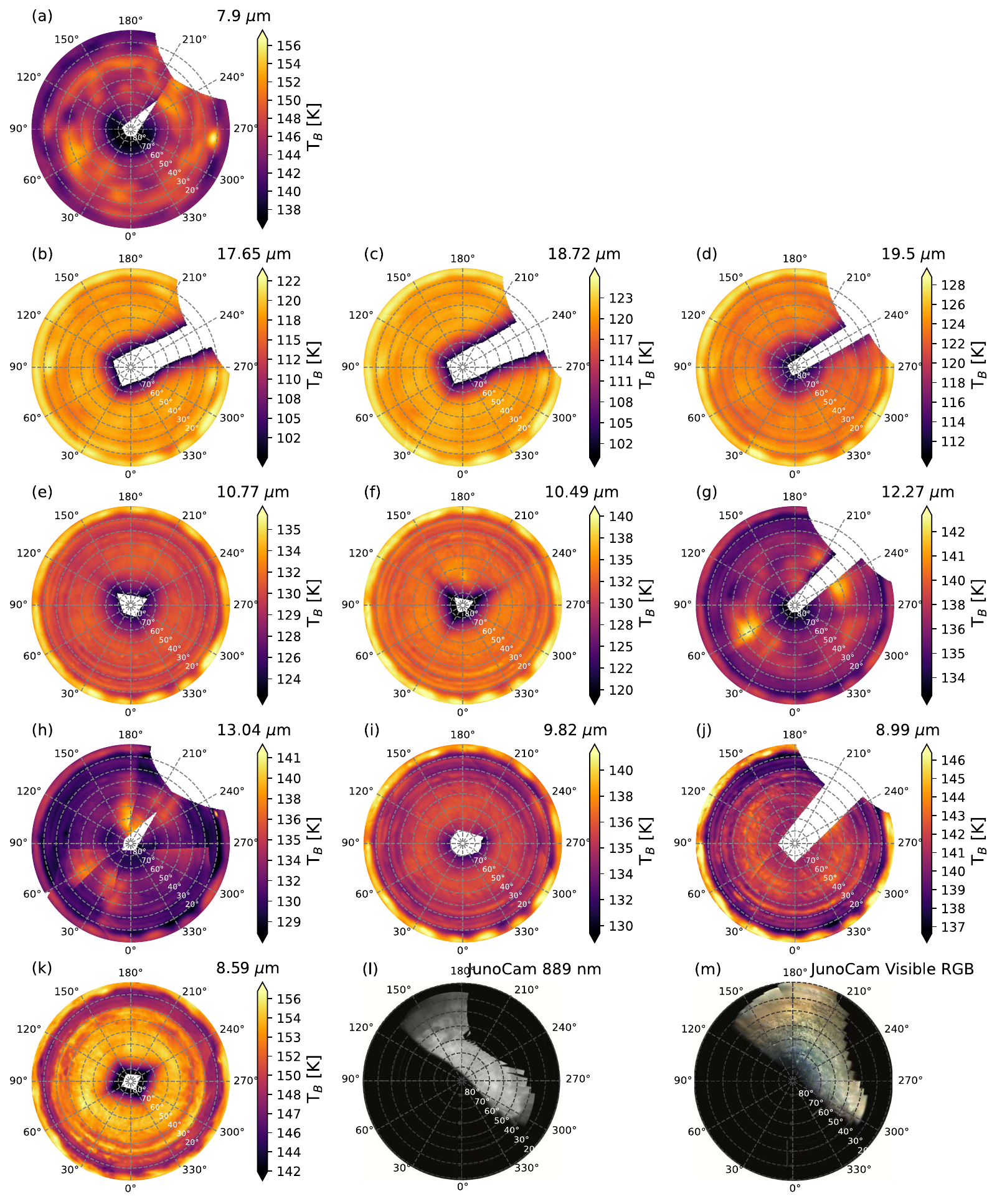}
    \caption{(a)-(k) Northern equidistant polar projections of the VLT/VISIR 2018 May 24$^{th}$-27$^{th}$ global radiance maps, corrected for limb brightening/darkening. At this epoch, the southern hemisphere of Jupiter was facing the Earth, hence the highest northern latitudes (poleward 80$^{\circ}$N) have been cut off in the northern polar projection due to extreme emission angle in this dataset and proximity to the edge of the disc (by 5"). Moreover, because of the interruption of the observations during the night of 24$^{th}$-25$^{th}$ due to strong winds, a large part of the planet between 210 to 270$^{\circ}$W System III is missing in these polar maps for seven of the eleven filters used here (7.9, 17.65, 18.72, 19.5, 12.27, 13.04 and 8.99~$\mu$m filters).
     (l) JunoCam CH$_4$-band (889 nm) and (m) visible RGB observations of Jupiter's north pole during the Perijove 13, processed by Gerald Eichstädt and John Rogers \cite{Roge:22}.  Longitude coordinates are displayed in the System III W longitude.}
    \label{fig:north_pole}
\end{figure}

\begin{figure}
    \centering
    \includegraphics[width=\textwidth]{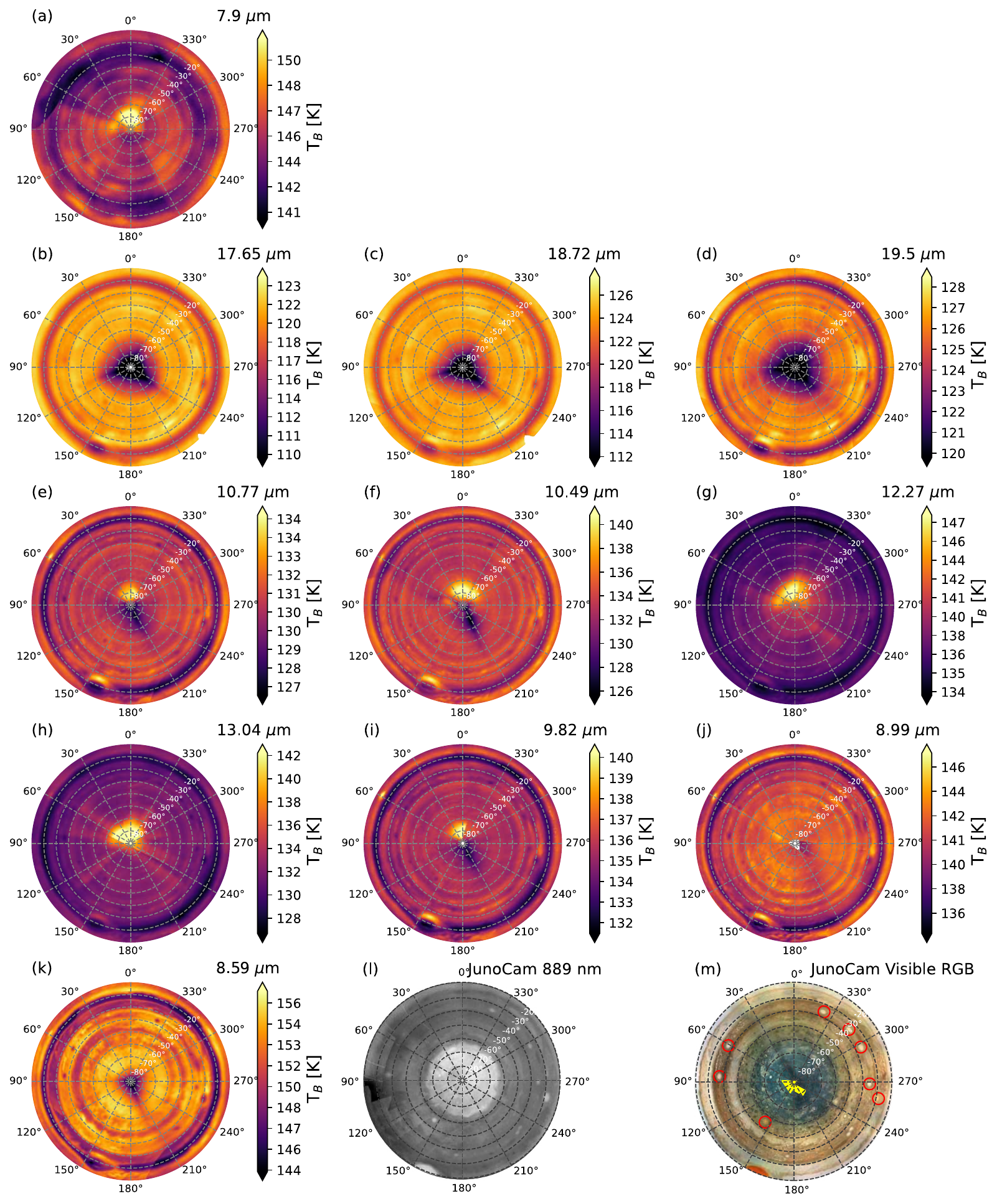}
    \caption{(a)-(k) Southern equidistant polar projection of the VLT/VISIR 2018 May 24$^{th}$-27$^{th}$ global maps, corrected for limb brightening/darkening. 
    (l) JunoCam CH$_4$-band (889 nm) and (m) visible RGB observations of Jupiter's north pole during the Perijove 13, processed by Gerald Eichstädt and John Rogers \cite{Roge:22}. Longitude coordinates are displayed in the System III W longitude.}
    \label{fig:south_pole}
\end{figure}

The zonal organisation of the high latitudes is different between the two hemispheres, as seen in both Figures \ref{fig:north_pole} and \ref{fig:south_pole}. 
These polar-projection maps show a very structured southern hemisphere with the characteristic banded structure persisting up to the southern jet S6 (following the nomenclature of \citeA{Roge:22}) at 64$^{\circ}$S in the upper troposphere and stratosphere. 
In comparison, the northern pole begins to show similar small-scale structure beyond the westward jet located at 68$^{\circ}$N at the tropopause (80 mbar, in Figure \ref{fig:north_pole}b), and exhibits intense wave activity that dominates in the stratosphere (Figure \ref{fig:north_pole}a).

The view of the southern pole is dominated by the southern aurora, seen primarily in the filters associated with emission lines of 
CH$_4$ (7.90 $\mu$m, indicative of warming in Figure \ref{fig:south_pole}a), 
C$_2$H$_4$ (10.77, 10.49 and 9.82 $\mu$m in Figures \ref{fig:south_pole}e, \ref{fig:south_pole}f and \ref{fig:south_pole}i), 
C$_2$H$_2$ (13.04 $\mu$m in Figure \ref{fig:south_pole}h), 
and C$_2$H$_6$ (12.27 and 13.04 $\mu$m in Figures \ref{fig:south_pole}g and \ref{fig:south_pole}h). 
The increased brightness temperature for 10.77, 10.49, 12.27, 13.04 and 9.82-$\mu$m filters can either be explained by enhanced heating, or potentially indicate an enhancement of hydrocarbon production from the ion-neutral chemical breakdown of CH$_4$. 
Considering the north pole, there is a slight increase of the mid-infrared emission in the 7.90, 12.27 and 13.04-$\mu$m filters between 120$^{\circ}$W System III and 210$^{\circ}$W System III, and poleward the 70$^{\circ}$N parallel (Figures \ref{fig:north_pole}a, \ref{fig:north_pole}g and \ref{fig:north_pole}h).
Nonetheless, we do not consider this to represent the presence of auroral heating in the northern hemisphere, since the area is wider, and the heat signature is more diffuse and not as bright as in the southern hemisphere.  
The northern auroral emission is known to change over short timescales \cite{Sinc:19CH4_aurora_solar_wind_compression}, and the May 2018 observations may have caught the north pole during a period when auroral heating was absent.

Moreover, this missing northern aurora signature enables a clear view of the northern cold polar vortex \cite{Orto:02, Porc:03, Kund:04, Orto:08, Flet:16jupiter}.
The low temperatures over both poles in the upper troposphere and stratosphere show that the cooling persists from 1 mbar (Figures \ref{fig:north_pole}h and \ref{fig:south_pole}h) to $\sim$5 mbar (Figures \ref{fig:north_pole}a and \ref{fig:south_pole}a) consistent with the aerosol cooling effect due to the large optical thickness and the fractal-aggregate shapes of the polar aerosols (one to two orders of magnitude optically thicker than the lower latitude haze), as described by \citeA{Zhan:13}.  
To investigate the relation between the cold polar vortex and Jupiter's polar aerosols, we compare to JunoCam observations of the polar region acquired during the VLT/VISIR run (perijove 13, 2018 May 25$^{th}$).

The general appearance of the poles in the JunoCam visible observations is bluish with discrete white features \cite{Orto:17PoleJup,Roge:22}, co-located with bright structures in the CH$_4$ band, implying that these reflective storm systems are located at higher altitudes than the surrounding less reflective regions (Figure \ref{fig:north_pole}m and \ref{fig:south_pole}m). 
This transition boundary to the polar region is seen in the JunoCam near-IR (Figures \ref{fig:north_pole}l and \ref{fig:south_pole}l) as a large CH$_4$-bright region extending poleward of 50-60$^{\circ}$N (with a ragged boundary in the north) and poleward of $\sim$64$^{\circ}$S (more discrete boundary in the south). 
This polar boundary meanders with longitude (possibly modulated by the presence of a Rossby wave, \citeA{Li:04, BarrIzag:08, Roge:22}) in JunoCam visible observations, which is not captured in the VLT/VISIR observations.

However, despite the limited 889-nm view of the north polar aerosols from JunoCam (Figure \ref{fig:north_pole}l), there does not seem to be a strong correlation with the northern cold polar vortex boundary and the hazes.  
Indeed, the northern cold polar vortex appears to be more confined to higher latitudes (Figures \ref{fig:north_pole}b and \ref{fig:north_pole}d). 
The correlation between the JunoCam aerosols and the cold polar vortex is much clearer in the south (at $\sim$67$^{\circ}$S, Figure \ref{fig:south_pole}), where the tropospheric cooling is colocated with the reflective aerosols at 889 nm, and with the dark blue polar domain in the RGB image. 
This strong correlation in the south, and weak correlation in the north, could potentially be explained by an asymmetry in the aerosol populations at the two poles \cite{Zhan:13, Guer:20}. 
If the aerosols are generated via ion-neutral chemistry within the auroral ovals, the fact that the south cold polar vortex completely encircles the compact south polar auroral oval means that those aerosols could be constrained by the vortex boundary acting as a transport barrier.  
Conversely, the northern auroral oval extends outside of the cold polar vortex, meaning that polar aerosols would not be entrained as efficiently by the vortex.  
This asymmetry in polar aerosols was also observed by \citeA{Zhan:13} in their study of aerosols derived from Cassini observations of Jupiter.

The radiative contribution of Jupiter's polar aerosols was discussed by \citeA{Guer:20} using a 1-D seasonal radiative-convective model, confirming the large contribution of the stratospheric polar aerosol in the radiative budget. 
Depending of the optical properties of these aerosols (which remain uncertain), the authors showed that the maximum net impact of stratospheric aerosol is located between 5 and 30 mbar at 45-60$^{\circ}$.
In their model, they also show a significant cooling poleward 65$^{\circ}$, consistent with the present VISIR dataset.

\subsection{Mid/High-Latitude Vortices}
\label{polar_vortices_storms}
The polar projections also provide a means for identifying discrete features at Jupiter's mid-latitudes that may be classified by their long-lived structure, features like Oval BA ($\sim$30$^{\circ}$S), the Anticyclonic White Ovals (AWOs, 35-37$^{\circ}$S) and S4-LRS-1 ($\sim$56$^{\circ}$S), using the classical nomenclature \citeA{Roge:95}. 
All of these anticyclones are associated with low brightness temperatures in VLT/VISIR 19.5, 10.77 and 8.59$\mu$m filters, suggestive of cold-core features above the clouds (Figures \ref{fig:south_pole}d, \ref{fig:south_pole}e and \ref{fig:south_pole}k; also seen in 9.82, 8.99$\mu$m filters in Figures \ref{fig:south_pole}i, \ref{fig:south_pole}j).  
They appear as white or reddish features in JunoCam visible and CH$_4$-band observations (Figures \ref{fig:south_pole}m, denoted by red circles, and \ref{fig:south_pole}l), suggesting that white aerosols (ices) are condensing in the cold upper parts of these anticyclones. 
The sensing through several wavelengths both in visible and infrared wavelengths argue that the structure of these anticyclones persists with height, with subtle changes of shape. 
Indeed, at the 500-750 mbar cloud-top level, the VLT/VISIR observations at 8.99$\mu$m exhibit very low radiance, correlated with the location of the thick cloud cover at 1-4 bar seen by \citeA{dePa:11}, with a bright peripheral ring of negligible cloud opacity as well. 
The associated cooler temperatures of this specific aerosol structure extend all the way to the tropopause (Figure \ref{fig:south_pole}b, corresponding to the 17.65-$\mu$m filter). 
As well as the anticyclonic ovals, the maps show a pale quiescent circulation just North-East of it at $\sim$50-70$^{\circ}$W (the ``South Temperate Belt Spectre''). 
This is dark at 0.89-$\mu$m (Figure \ref{fig:south_pole}l) and warm at 8.99-10.77 $\mu$m (Figures \ref{fig:south_pole}j and \ref{fig:south_pole}e, respectively).  

These mid-latitude features provide some insight into the smaller white ovals observed in the polar regions by JunoCam. 
Some of these smaller white ovals (between 70 to 80$^{\circ}$S) exhibit spiral extensions (seen in the Juno/JunoCam observations in Figure \ref{fig:south_pole}m) that indicates their anticyclonic motion, although many lack this structure, giving little indication of their direction of rotation. 
Some of those that display anticyclonic motion are co-located with cloudy regions at 2-4 bar, implying similar overall structure to the larger white ovals at mid-latitudes, albeit on a smaller scale. 
The VLT/VISIR maps suggest that the vertical structure of these AWOs is similar in both hemispheres.

Apart from the anticyclones, the polar regions of both hemispheres exhibit amorphous cyclonic structures referred to as ``Folded Filamentary Regions'' (FFR, using the nomenclature of \citeA{Inge:84}, one of them is pointed by the yellow arrow on \ref{fig:south_pole}m). 
These regions are among the brightest and largest discrete features in the polar regions and have been previously observed by Voyager, Cassini, Hubble and Juno \cite{Porc:03, Li:04, BarrIzag:08, Roge:22}.
In the south, several FFRs are located in a band between 65$^{\circ}$S and 72$^{\circ}$S (\citeA{Roge:22}, Figure \ref{fig:south_pole}). 
These regions of chaotic flow, like the AWOs, are highly reflective in the methane band, implying elevated cloud tops, although probably lower than the polar hoods, which may represent enhancement of lower stratospheric aerosols consistent with the expected distribution of aerosols at the poles \cite{Zhan:13}). 
The FFRs have associated structure at depth, deep cloud opacity that is modulated by their cyclonic motion \cite{Inge:84}. 
Identification of discrete structures beyond 70$^{\circ}$S in the other tropospheric mid-IR filters is challenging due to the proximity to the limb and the general cold temperatures of the vortex in which they are embedded. 
Nevertheless, the FFRs generally appear to be warmer than their surroundings and cloud-free in the mid-IR, consistent with their identification as cyclonic, stormy regions in visible imaging.  
Even with the excellent spatial resolution afforded by VLT/VISIR, diagnosing the thermal conditions within these high-latitude features remains extremely difficult without mid-IR capabilities on an orbiting spacecraft.

\section{Temporal Changes in the Southern Aurora}
\label{sec:polar_aurora_retrieval}

\begin{figure}
    \centering
    \includegraphics[width=\textwidth]{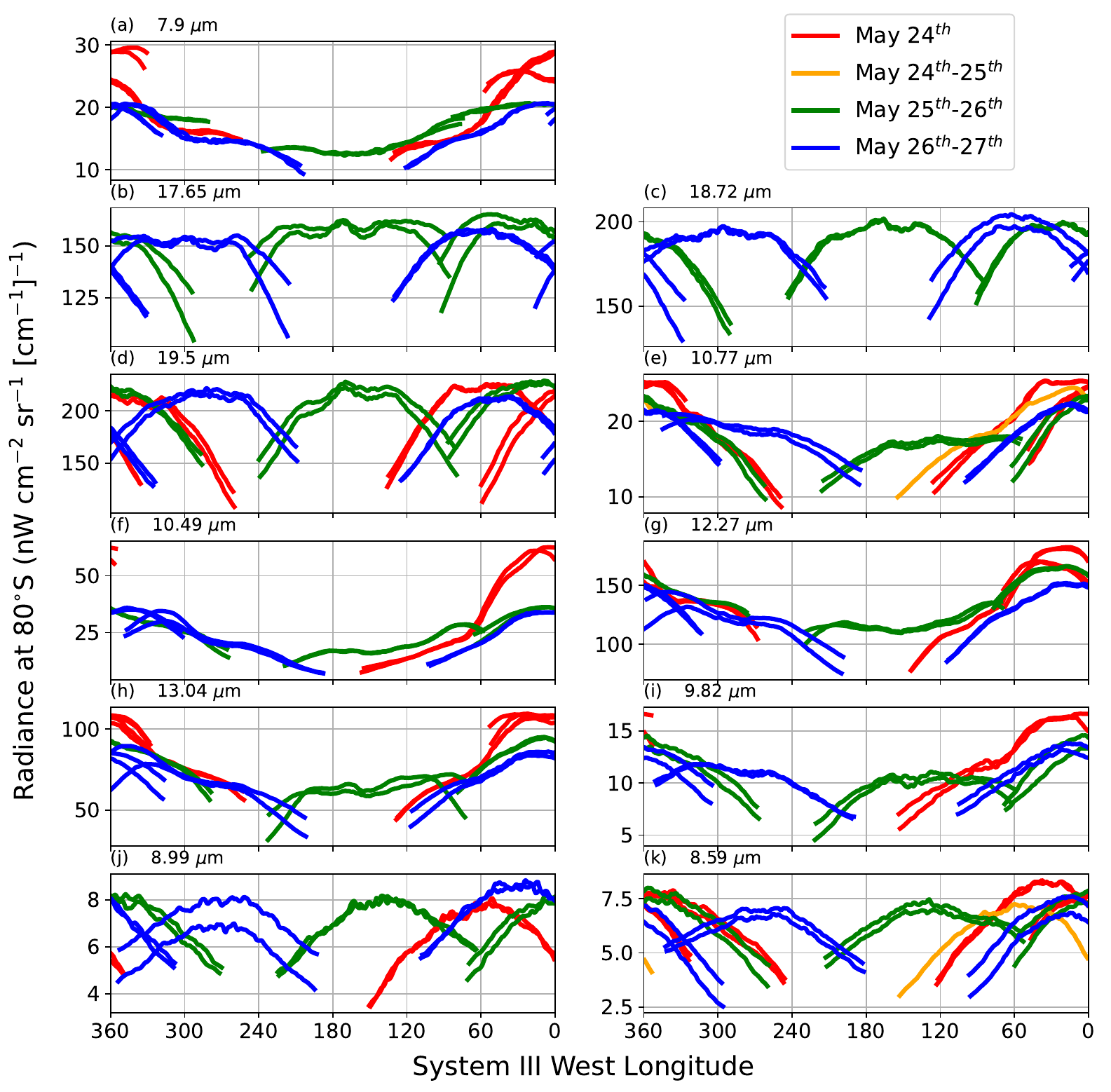}
    \caption{Night-colored-coded longitudinal radiance profiles along the 80$^{\circ}$S parallel for the entire dataset retaining the limb-darkening/brightening effect for VLT/VISIR N-Bands (7.9 and 10.77 to 8.59$\mu$m) and Q-Bands (19.50, 18.72 and 17.65 $\mu$m).}
    \label{fig:radiance_profiles_parallel80S}
\end{figure}

Along the 80$^{\circ}$S parallel (Figure \ref{fig:radiance_profiles_parallel80S}), there is a large temporal evolution of the observed radiance over the southern auroral region (0$^{\circ}$--60$^{\circ}$ and 300$^{\circ}$--360$^{\circ}$ System III West longitude), that is also inferred in the zonally averaged profiles of radiance (Figure \ref{fig:calib_profile}). 
This time evolution is mainly observed in the methane emission band filter 7.9~$\mu$m, ethylene (C$_2$H$_4$) and acetylene (C$_2$H$_2$) emission band filters (10.49, 10.77 and 13.04~$\mu$m), and ethane (C$_2$H$_6$) emission band filters 12.27~$\mu$m, as well as the 9.82~$\mu$m filter (Figures \ref{fig:radiance_profiles_parallel80S}a, \ref{fig:radiance_profiles_parallel80S}e, \ref{fig:radiance_profiles_parallel80S}f, \ref{fig:radiance_profiles_parallel80S}g, \ref{fig:radiance_profiles_parallel80S}h, \ref{fig:radiance_profiles_parallel80S}i, respectively). 
Brightness variability of the aurora is so large for the 7.9, 10.49, 13.04 and 9.82~$\mu$m filters from one night to the next that this produces discontinuities along the 80$^{\circ}$S parallel profiles of radiance for these filters. 
In particular, on the edges of the aurora location such as $\sim$320$^{\circ}$W (Figures \ref{fig:radiance_profiles_parallel80S}f and \ref{fig:radiance_profiles_parallel80S}i), 300$^{\circ}$W (Figure \ref{fig:radiance_profiles_parallel80S}a) and 90$^{\circ}$W (Figure \ref{fig:radiance_profiles_parallel80S}h). 

To follow the average time variation of the thermal aurora signature in this mid-infrared dataset, we have gathered the images per night of observation (i.e., from day i$^{th}$, 11p.m. to day (i+1)$^{th}$, 7a.m.). 
For each VLT/VISIR filter, we have averaged the data from 0$^{\circ}$ to 60$^{\circ}$ and from 300$^{\circ}$ to 360$^{\circ}$ longitude System III West along a 2$^{\circ}$-latitude band centered at 80$^{\circ}$S.
These two data bins correspond to the maximum emission of the aurora as depicted in the zonal profiles of radiance in Figure \ref{fig:radiance_profiles_parallel80S}.  
This regional averaging creates a single spectral point per filter to characterise the average atmospheric response of the aurora event during the four observing nights.
However, the second night of observation, from May 24$^{th}$ 11pm to May 25$^{th}$ 7 am, had to be shortened due to bad weather, and therefore only two filters constitute this nightly dataset (8.59 and 10.77 $\mu$m filters). 
In addition, the observing night May 24$^{th}$ from 1 am to 7 am is also a reduced dataset of 9 filters (7.9, 19.5, 10.49, 10.77, 12.27, 13.04, 9.82, 8.99 and 8.59~$\mu$m).
So, to keep retrieval calculation consistent from one night to the next, we have only considered the first, third and fourth nights of observation (i.e., May 24$^{th}$, May 25$^{th}$-26$^{th}$ and May 26$^{th}$-27$^{th}$ in what follows), with a spectral dataset only composed of the 9 shared filters (i.e., excluding the 17.65 and 18.72~$\mu$m filters).

\begin{figure}
    \centering
    \includegraphics[width=0.9\textwidth]{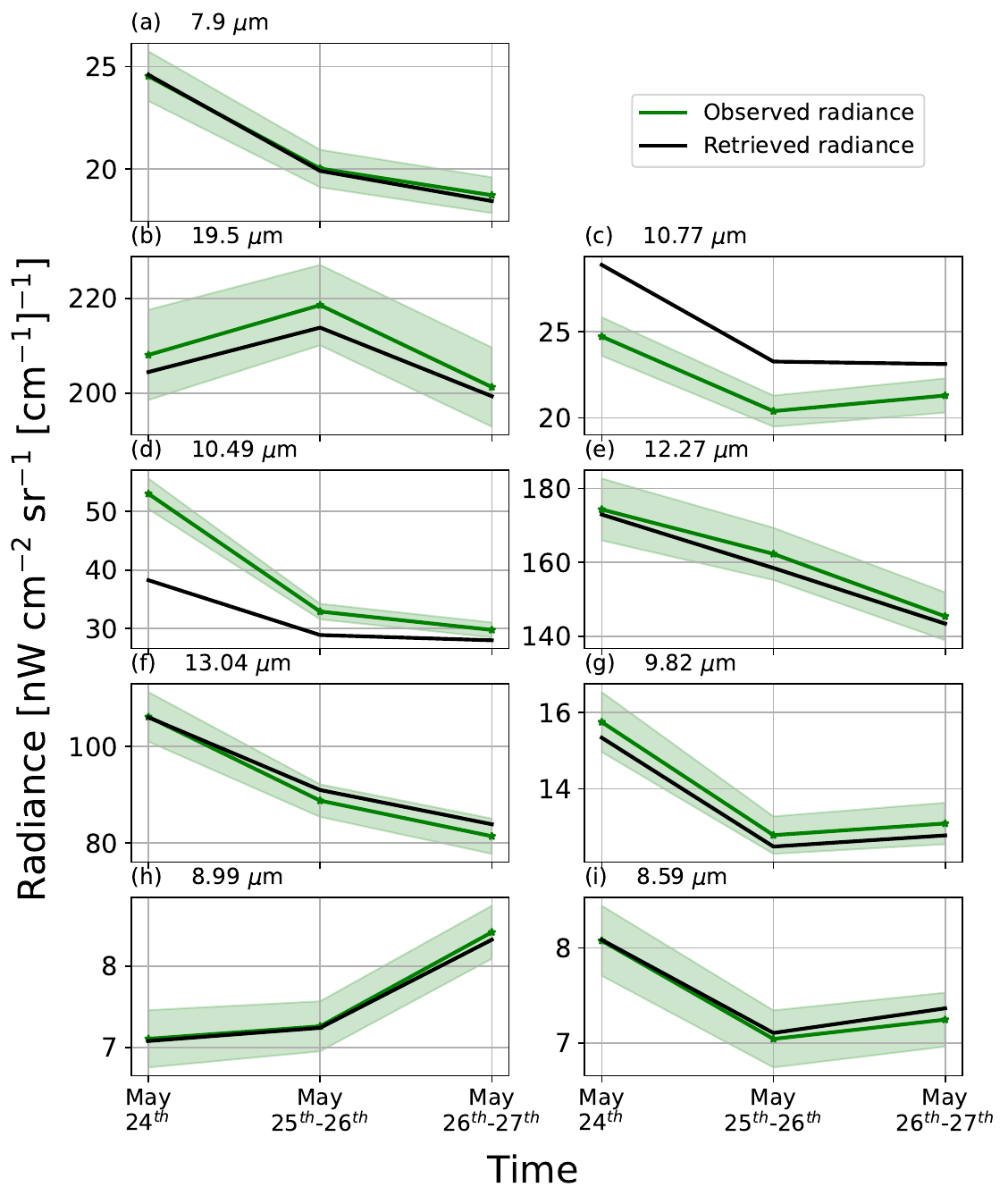}
    \caption{Observed and retrieved radiance time profiles for the averaged aurora area during the three selected nights.}
    \label{fig:aurora_radiances_over_time}
\end{figure}

For this particular retrieval, the reference atmosphere is composed by the same prior as the zonal mean study above (see section \ref{sec:retrieval}).

The observed radiance from the 24$^{th}$ to the 27$^{th}$ of May decreased with time for most of the wavelengths sensing the temperature and hydrocarbons species in the stratosphere (Figure \ref{fig:aurora_radiances_over_time}).
Most of the retrieved radiance profiles for the nine selected wavelengths are within the 5\% error envelope and so are quite well fitted, in particular the stratospheric temperature 7.9-$\mu$m filter, the tropospheric aerosol opacity 8.99 and 8.59-$\mu$m filters. 
Only the 12.27 and 13.04-$\mu$m filters, sensing the stratospheric ethane and acetylene, as well as the 10.49 and 10.77-$\mu$m filters, sensing the stratospheric high latitude ethylene distribution, are not well retrieved, even with the addition of the scaling of C$_2$H$_4$ profile for the retrieval calculations. 
Considering the high emission angle of those data, this radiance fitting is the best we could obtain.

\begin{figure}
    \centering
    \includegraphics[width=0.6\textwidth]{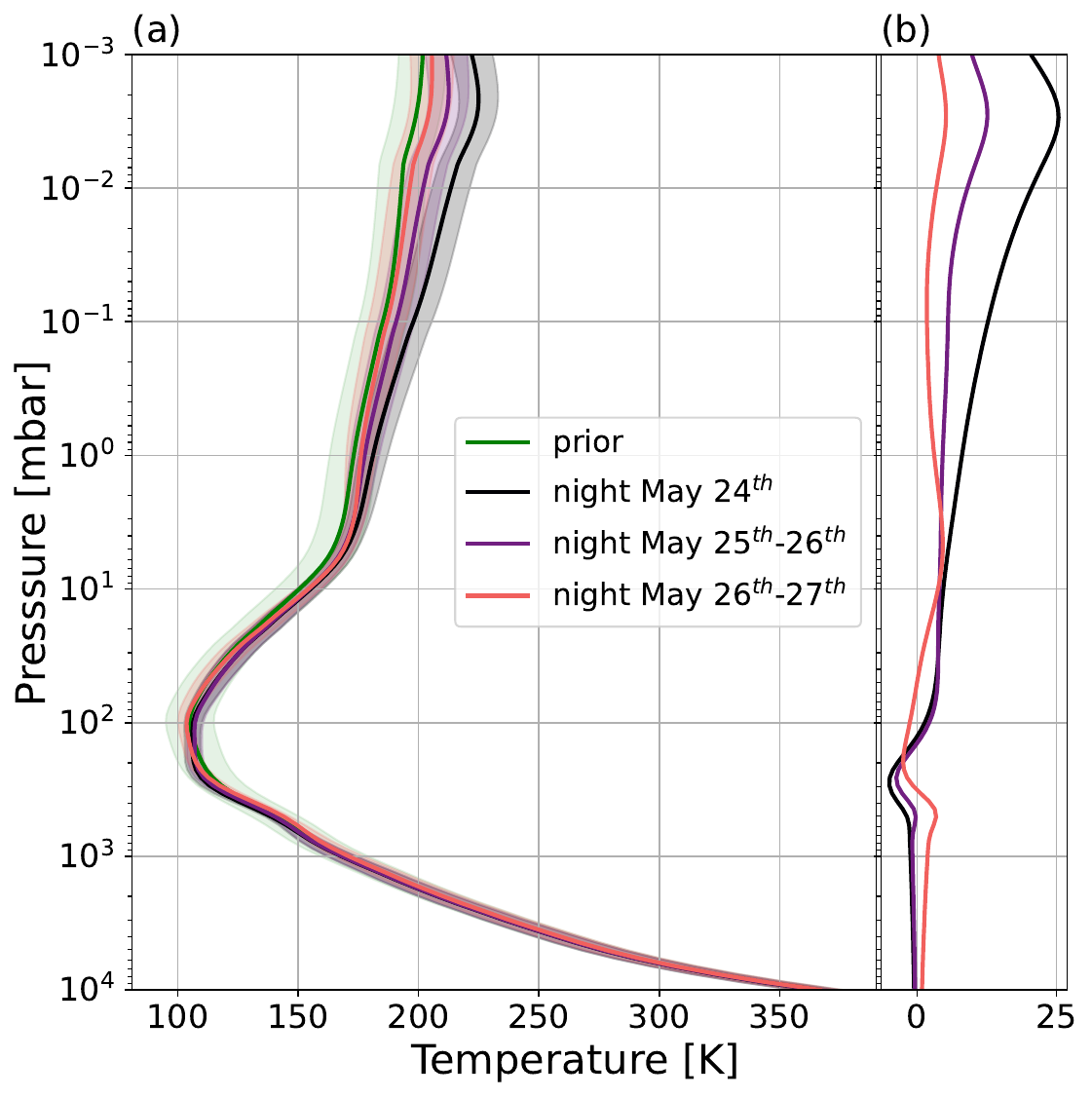}
    \caption{(a) Retrieved temperature profiles and (b) deviation from the prior profiles ($T_{retrieved} - T_{prior}$) over the aurora area for each of the three selected nights.}
    \label{fig:aurora_temperature_profiles_over_time}
\end{figure}

The retrieved temperature profiles for the three nights between the 24$^{th}$ and 27$^{th}$ of May 2018 show a significant cooling, by almost 20~K, in the stratosphere over time (Figure \ref{fig:aurora_temperature_profiles_over_time}). 
If we assume this cooling is just the effect of temperature variation, it implies a cooling times of a few hours that are unrealistically short at these pressures. 
So it is therefore more likely that some kind of direct non-LTE (local thermal equilibrium) effects are adding to the regular LTE emission, allowing these things to appear to change on very short timescales.  
That would imply the VLT/VISIR observation is not truly measuring the temperatures of these regions. 
Regarding the contribution function of each VLT/VISIR filters at 80$^{\circ}$S (Figure \ref{fig:weighting_functions}), it is challenging to determine which chemical species are responsible for the stratospheric cooling. 
Emission from the dominant hydrocarbons, ethane and acetylene, and to a lesser extent methane and ethylene, could result in radiative time constants that are as short as a few hours at the very highest altitudes over the poles, allowing the neutral atmosphere to cool down following an injection of auroral energy over a short timespan.  
Brightening and dimming of the southern auroral emission was observed over comparable timescales in 2017 by Subaru/COMICS \citeA{Sinc:19CH4_aurora_solar_wind_compression}.

\begin{figure}
    \centering
    \includegraphics[width=0.6\textwidth]{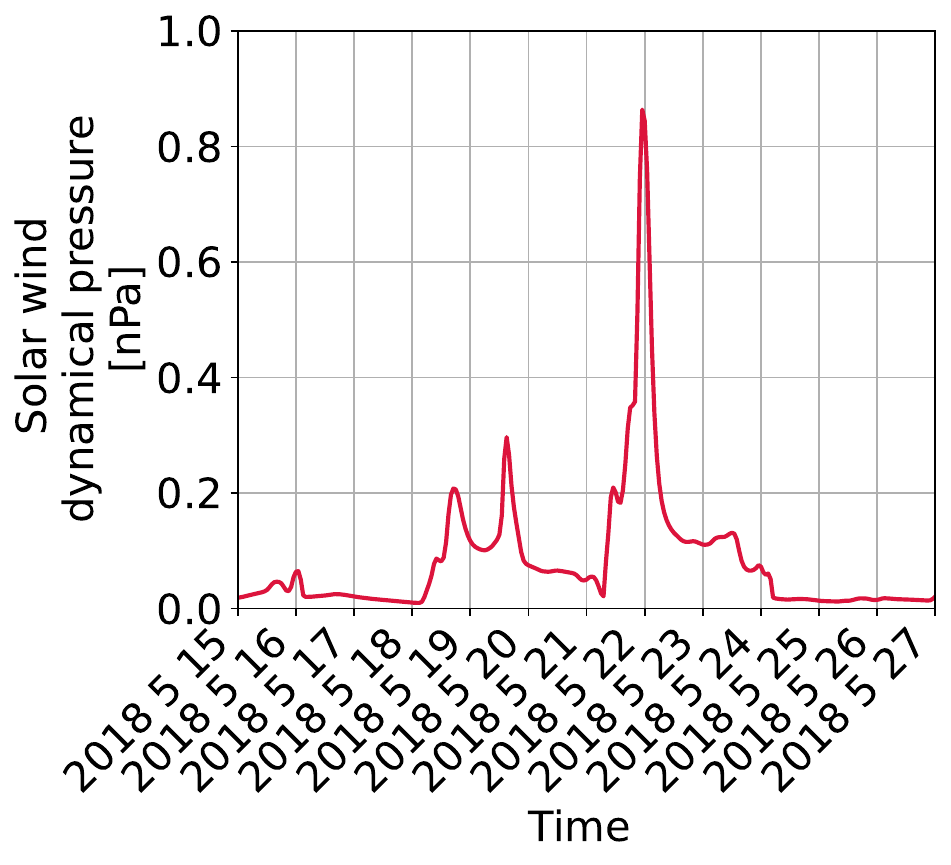}
    \caption{Predicted solar wind conditions in 2018, calculated from the OMNI-measured conditions at Earth, and the 1-D model by \citeA{Tao:05} to propagate and predict the solar wind flow at Jupiter.}
    \label{fig:solar_wind_TAO_model}
\end{figure}
The solar wind activity predictions of a uni-dimensional idealised magnetohydrodynamic model for Jupiter \cite{Tao:05} have been analysed to look for auroral events during May 2018.
Calculated in a heliocentric frame, with measurement of the solar wind flow around the Earth as inputs, the prediction of the model is at its most accurate when the target planet is close to opposition, which for Jupiter was on May 8$^{th}$ 2018 (Earth-Sun-Jupiter angle).
On May 23$^{rd}$ 2018, Jupiter was describing, with the OMNI spacecraft, an angle of -2$^{\circ}$ on the heliocentric longitude frame, which implies a maximum of prediction accuracy \cite{Zieg:08}.  
Between the 20$^{th}$ and the 21$^{st}$ of May 2018, the model predicts an enhancement of the solar wind dynamical pressure by 0.85~nPa in the jovian environment, followed by a period of quiescent conditions (Figure \ref{fig:solar_wind_TAO_model}). 
This solar wind event may have perturbed the magnetosphere, transiently heating the upper atmosphere through higher rates of particle precipitation and/or Joule heating driven by stronger Pedersen currents  \cite{Grod:01,Yate:14,Badm:15}.   
Even with the timing uncertainty, the VLT/VISIR observations from May 24$^{th}$ to the 27$^{th}$ may have captured the subsequent cooling of the southern auroral emission after this solar wind compression event.  
It is unclear why the northern auroral oval did not brighten in response to the same compression event, hinting that not all compression responses are equal.

\newpage
\section{Conclusions}
\label{sec:conclusion}

Global pole-to-pole ground-based mid-infrared VLT/VISIR maps of Jupiter, spanning 5-20 $\mu$m, were obtained on the Very Large Telescope on 2018 May 24$^{th}$-27$^{th}$.  These represent the highest quality mid-infrared maps acquired by VISIR during a programme of ground-based support for NASA's Juno mission since 2016, and thus they provide our best coverage of the cold polar vortices and auroral activity.  A pipeline for destriping, cleaning, calibrating and projecting the data was designed, and applied to all data of 2018 May 24$^{th}$ to 27$^{th}$ (raw data are available on the ESO Science Archive Facilities website \url{http://archive.eso.org/eso/eso_archive_main.html}, and the processed data are available to the community in our open-access archive \cite{bardet_deborah_2023_8401816}), with findings spanning multiple datasets to be detailed in forthcoming studies.  From this map in 2018, we draw the following conclusions:

\begin{itemize}
    \item \textbf{Belts and Zones: } Jupiter's banded structure of cool zones and warm belts extends from the equator to the edge of the polar vortices, with a strong correlation between the maximum vertical windshear (from temperature gradients) and the locations of Jupiter's zonal jets.  The zonal jets decay throughout the upper tropospheric levels (80-600 mbar) sensed by the VISIR data.  Peaks in aerosol opacity sometimes occur within the cool zones (suggestive of condensation of saturated vapours), but the correlation between aerosol opacity and the zonal jet structure is weaker, confirming that aerosols are not a good proxy for Jupiter's belts and zones.
    \item \textbf{Equatorial Oscillation:}  A warm equatorial peak near 1-4 mbar, and associated off-equatorial minima in stratospheric temperatures, reveal the presence of Jupiter's equatorial stratospheric oscillation (sometimes called the QQO).  The presence of the warm equatorial band in 7.9 $\mu$m imaging in 2018 is consistent with the downward propagation of the wave pattern shown by \citeA{Gile:20} in the year after a 2017 disruption.   
    \item \textbf{Hydrocarbon Distributions:}  Although Jupiter's low latitudes can be fitted with a horizontally uniform distribution of ethane and acetylene, this is not true at high latitudes, where enhancements of both ethane and acetylene are required (these hydrocarbons have more significant contributions at the higher emission angles). Hence, as already suggested by previous studies, these enhancements of hydrocarbons reinforce the evidence of enhanced ion-related chemistry in Jupiter's auroral regions. 
    \item \textbf{Cold Polar Vortices:}  The banded structure of low/mid-latitudes gives way to cold polar vortices at both poles in the upper troposphere, and over the north pole in the stratosphere, for whose the latitudinal boundaries are located at 67$^{\circ}$S and 64$^{\circ}$N, but the southern stratosphere is somewhat masked by auroral emission (see below).  The southern polar vortex coincides with reflective aerosols observed in the CH$_4$ band at 890 nm by JunoCam, and with the blue colouration in visible light, suggesting polar aerosols entrained by the vortex.  In the north, the aerosol field is more diffuse and not as well correlated with the edges of the cold polar vortex.  The cold vortices are likely due to radiative cooling associated with these aerosols, entrained dynamically by the northernmost and southernmost jet streams.
    \item \textbf{Auroral Heating:}  The southern auroral oval was bright in the mid-infrared in May 2018 (although the northern oval was less visible), where the influx of energetic electrons and ions from the jovian magnetosphere and external solar-wind environment cause warming in the stratosphere. A solar wind compression is expected to have reached Jupiter immediately before the VISIR observations, and over four consecutive nights of observations VISIR captured the subsequent cooling of the southern auroral region, from the 24$^{th}$ to the 27$^{th}$ of May 2018.  Despite this evidence of a interaction between the magnetosphere and stratosphere in Jupiter's polar region, there is currently no clear picture of how the particle precipitation inside the aurora oval interacts with the lower atmosphere, in terms of chemical and dynamical processes. 
    \item \textbf{Anticyclones and Cyclones:} The high spatial resolution provided by VLT/VISIR resolves vortices of many different sizes, from the small white ovals and folded filamentary regions at high latitudes, to larger white anticyclones at mid-latitudes, to Oval BA and the Great Red Spot.  A prominent brown barge is observed to be bright in the SEB, indicating warm, cloud-free and volatile-depleted conditions within these cyclones.  Mid-infrared observations probe altitudes above vortex midplanes in the deeper troposphere, so anticyclones are cold, and cyclones warm, consistent with the decay of their tangential velocities with altitude.  Detailed characterisation of the GRS confirms previous studies, with a small warm core embedded within the larger cold vortex, and complex rifting in the surroundings.  The GRS may influence the stratospheric temperatures near 5 mbar, with warm patches immediately northeast and northwest of the vortex, possibly associated with upward-propagating wave activity.
\end{itemize}

The May 2018 VISIR observations provide only a snapshot of Jupiter's atmosphere, but demonstrate a wealth of phenomena that could be monitored over time using the 2016-2022 VISIR time series. 
To further understand the links between the auroral regions, the chemistry of the surrounding polar atmosphere and the mid-latitude regions, from the magnetosphere to the troposphere, we require longitudinally-resolved and time variable ion-neutral photochemical model of Jupiter's high latitudes, coupled with a global climate model.
Although coupled chemistry and transport models do exist, the inclusion of auroral contributions, both of the hydrocarbons and of the aerosols responsible for radiative cooling, is needed to improve our knowledge of Jupiter's polar atmosphere.

\section*{Acknowledgments}
Bardet, Fletcher, Antuñano, Roman and Melin were supported by a European Research Council Consolidator Grant (under the European Union’s Horizon 2020 research and innovation program, grant agreement No 723890) at the University of Leicester.  
Donnelly acknowledges funding from Agence Nationale de la Recherche (ANR), project EMERGIANT ANR-17-CE31-0007.  
Some of this research was carried out at the Jet Propulsion Laboratory, California Institute of Technology, under a contract with the National Aeronautics and Space Administration (80NM0018D0004).
This research used the ALICE High Performance Computing Facility at the University of Leicester.   
This paper is based on observations collected at the European Organisation for Astronomical Research in the Southern Hemisphere under ESO programme 0101.C-0073(A).  
We wish to thank the many ESO telescope operators and support staff who work long nights to provide the community with the data needed to better understand Jupiter.  

For the purpose of Open Access, the corresponding author has applied a CC-BY public copyright licence to any Author Accepted Manuscript (AAM) version arising from this submission.

\section*{Open Research}
The NEMESIS radiative transfer and spectral retrieval tool \cite{Irwi:08} is open-access and is available for download \cite{patrick_irwin_2022_5816724}.  
The processed data of May 2018 observations are available here  \cite{bardet_deborah_2023_8401816}, and the Python program used to calibrate, map the data, as well as create the input files for NEMESIS and plot the retrieval results is available here \cite{deborah_bardet_2023_10053811}.

\bibliography{newfred,references}

\end{document}


%
%


\title{Supporting Information for "Investigating Thermal Contrasts Between Jupiter's Belts, Zones, and Polar Vortices with VLT/VISIR"}
%
%

%
%


\authors{
Deborah Bardet\affil{1}, 
Padraig T. Donnelly\affil{2}, 
Leigh N. Fletcher\affil{1}, 
Arrate Antuñano\affil{3}, 
Michael T. Roman\affil{1}, 
James A. Sinclair\affil{4}, 
Glenn S. Orton\affil{4}, 
Chihiro Tao\affil{5},
John H. Rogers\affil{6}, 
Henrik Melin\affil{1}, 
Jake Harkett\affil{1}}

\affiliation{1}{\emph{School of Physics and Astronomy, University of Leicester}, address: University Road, Leicester LE1 7RH, United Kingdom}
\affiliation{2}{\emph{Laboratoire de M\'{e}t\'{e}orologie Dynamique / Institut Pierre-Simon Laplace (LMD/IPSL), Sorbonne Universit\'{e}, Centre National de la Recherche Scientifique (CNRS), \'{E}cole Polytechnique, Institut Polytechnique de Paris, \'{E}cole Normale Sup\'{e}rieure (ENS), PSL Research University}, address: Campus Pierre et Marie Curie BC99, 4 place Jussieu, 75005 Paris, France}
\affiliation{3}{\emph{UPV/EHU, Escuela Ingernieria de Bilbao, Fisica Aplicada}, address: Spain}
\affiliation{4}{\emph{Jet Propulsion Laboratory/ California Institute of Technology}, address: 4800 Oak Grove Drive, Pasadena, CA 91109, USA}
\affiliation{5}{\emph{National Institute of Information and Communications Technology (NICT)}, address:
4-2-1, Nukui-Kitamachi, Koganei, Tokyo 184-8795, Japan}
\affiliation{6}{\emph{British Astronomical Association}, address: London, United Kingdom}

\noindent\textbf{Contents of this file}

\noindent\textbf{Additional Supporting Information (Files uploaded separately)}
\begin{enumerate}
\item Captions for large Tables S1 (upload as separate excel file "dataset-tables.xlsm")
\end{enumerate}

\noindent\textbf{Introduction}


%
%
%
%
%
%
%
%
%
\begin{table}
\settablenum{S1} 
\caption{Images list composing the VLT/VISIR 2018 May 24-27 dataset}
\end{table}